\documentclass{article}

\usepackage{amsmath,amsfonts,bm}

\def\eqref#1{equation~\ref{#1}}

\def\1{\bm{1}}

\DeclareMathAlphabet{\mathsfit}{\encodingdefault}{\sfdefault}{m}{sl}
\SetMathAlphabet{\mathsfit}{bold}{\encodingdefault}{\sfdefault}{bx}{n}

\DeclareMathOperator*{\argmin}{arg\,min}

\usepackage{microtype}
\usepackage{graphicx}
\usepackage{subfigure}
\usepackage{booktabs} 
\usepackage{hyperref}
\usepackage{multirow}
\usepackage{array}
\usepackage[table]{xcolor}
\usepackage[dvipsnames]{xcolor}
\usepackage{siunitx}
\usepackage{subfigure}

\usepackage[accepted]{icml2025}
\makeatletter
\renewcommand{\ICML@appearing}{}
\makeatother

\usepackage{algorithmic}
\usepackage{algorithm}
\usepackage{amsmath}
\usepackage{amssymb}
\usepackage{mathtools}
\usepackage{amsthm}
\usepackage{inconsolata}
\usepackage{times}
\usepackage{amsfonts}
\usepackage{subcaption}
\usepackage{threeparttable}
\usepackage[table,xcdraw]{xcolor}

\usepackage{wrapfig}
\usepackage{bbm}

\usepackage[capitalize,noabbrev]{cleveref}
\AddToHook{cmd/appendix/before}{\crefalias{section}{appendix}}

\theoremstyle{plain}

\theoremstyle{definition}

\theoremstyle{remark}

\usepackage{color}

\newcommand{\eat}[1]{} 

\eat{}

\newcommand{\sym}[1]{\ifmmode^{#1}\else\(^{#1}\)\fi}
\newcommand{\ouralgo}{Inf-DDS}
\usepackage[textsize=tiny]{todonotes}

\icmltitlerunning{Influence Guided Sampling for Domain Adaptation of Text Retrievers}

\begin{document}

\twocolumn[
\icmltitle{Influence Guided Sampling for Domain Adaptation of Text Retrievers}




\begin{icmlauthorlist}
\icmlauthor{Meet Doshi}{comp}
\icmlauthor{Vishwajeet Kumar}{comp}
\icmlauthor{Yulong Li}{comp}
\icmlauthor{Jaydeep Sen}{comp}
\end{icmlauthorlist}

\icmlaffiliation{comp}{IBM Research AI}

\icmlcorrespondingauthor{Meet Doshi}{meet@ibm.com}

\icmlkeywords{dynamic data sampling, text retrievers, domain adaptation}

\vskip 0.3in
]



\printAffiliationsAndNotice{}  

\begin{abstract}
General-purpose open-domain dense retrieval systems are usually trained with a large, eclectic mix of corpora and search tasks. How should these diverse corpora and tasks be sampled for training? Conventional approaches sample them uniformly, proportional to their instance population sizes, or depend on human-level expert supervision. It is well known that the training data sampling strategy can greatly impact model performance. However, how to find the optimal strategy has not been adequately studied in the context of embedding models. We propose \textbf{\ouralgo{}}, a novel reinforcement learning–driven sampling framework that adaptively reweighs training datasets guided by influence-based reward signals and is much more lightweight w.r.t. GPU consumption. Our technique iteratively refines the sampling policy, prioritizing datasets that maximize model performance on a target development set. We evaluate the efficacy of our sampling strategy on a wide range of text retrieval tasks, demonstrating strong improvements in retrieval performance and better adaptation compared to existing gradient-based sampling methods, while also being $1.5{\times}-4{\times}$ cheaper in GPU compute. Our sampling strategy achieves a \(\mathbf{5.03}\) absolute \textsc{ndcg@10} improvement while training a multilingual \href{https://huggingface.co/BAAI/bge-m3}{\emph{bge-m3}} model and an absolute \textsc{ndcg@10} improvement of \(\mathbf{0.94}\) while training \href{https://huggingface.co/sentence-transformers/all-MiniLM-L6-v2}{\emph{all-MiniLM-L6-v2}}, even when starting from expert-assigned weights on a large pool of training datasets.
\end{abstract}

\section{Introduction}
\label{sec:into}

\begin{figure}[!t]
  \centering
  \includegraphics[width=0.95\columnwidth]{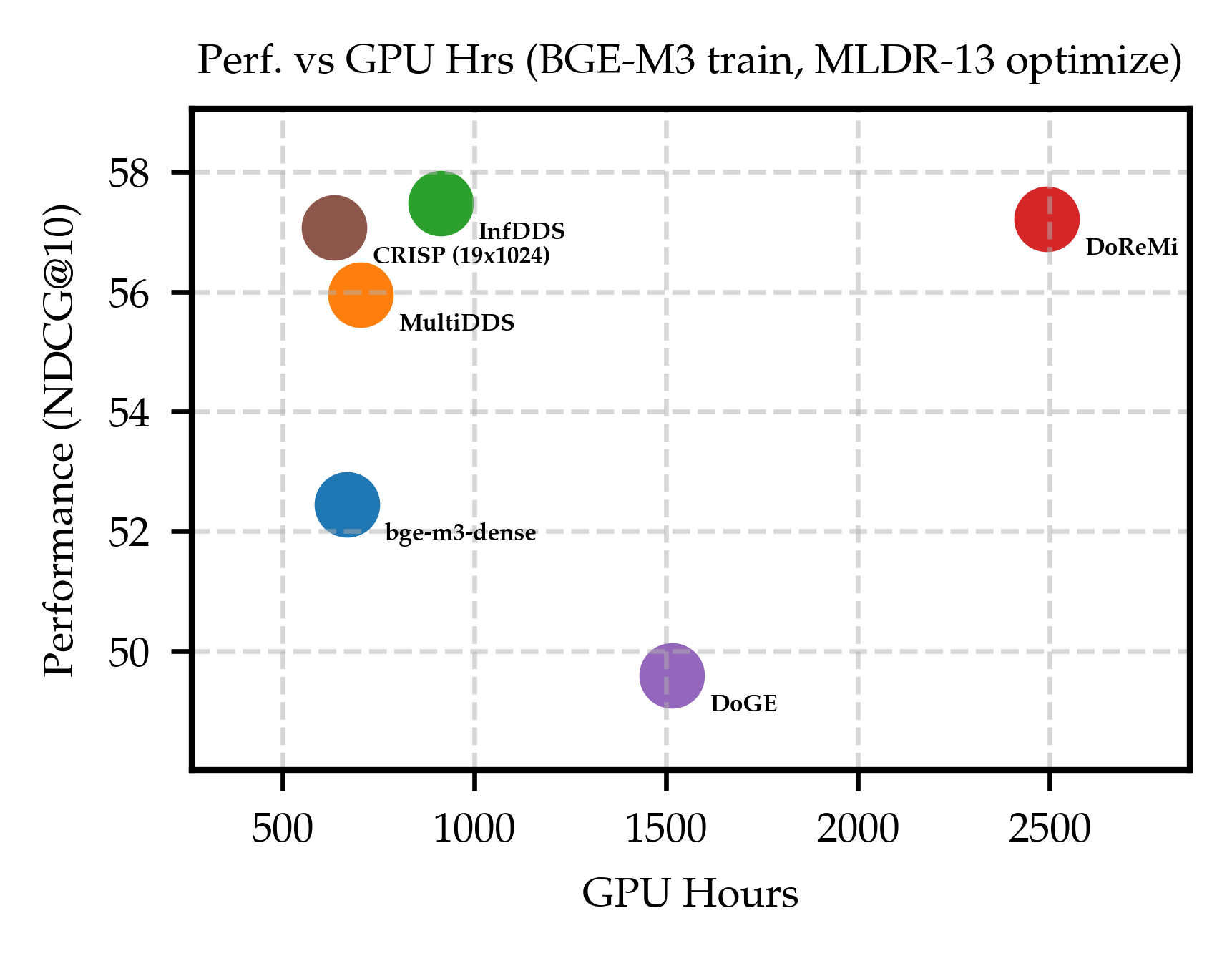}
  \caption{Trade-off between training time (GPU hours) and average NDCG@10 on the MLDR-13 test set when optimizing \emph{bge-m3-dense} on the MLDR-13 dev set.}
  \label{fig:bgem3-mldr13}
\end{figure}

Text-to-embedding based dense retriever models have recently gained huge popularity with their strong results across various benchmarks \citep{karpukhin-etal-2020-dense,DBLP:journals/tmlr/IzacardCHRBJG22, wang-etal-2023-simlm, bgem3, DBLP:journals/corr/abs-2404-05961}. These models are prized for their generalizability across domains without domain-specific tuning, typically trained on vast, diverse datasets. For instance, the \href{https://huggingface.co/sentence-transformers}{Sentence-Transformers} project, which develops generic sentence embeddings, utilized billions of instances across multiple datasets, with domain-specific data varying widely in size. However, larger datasets don’t inherently improve embedding quality, making it crucial to identify the most informative datasets and their optimal proportions for training. Effective sampling strategies are essential to prevent overfitting or underfitting, making dataset selection a central challenge in developing robust, generalizable, or domain-specific embedding models.

While random sampling is a common default, it is limited by ignoring data source informativeness when sampling from large training datasets. Alternatives include temperature sampling and instance-based proportional sampling. A more intensive approach involves creating ad-hoc sampling distributions via iterative experimentation, termed expert weights, requiring expert evaluation. However, these strategies are static and predefined, often suboptimal compared to the unknown ideal distribution for maximizing model performance. A dynamic sampling approach, capable of adaptation, may better approximate this optimal distribution.

There has consequently been substantial interest in making sampling adaptive. Gradient-based approaches such as DDS \citep{DDS} and its multi-target extension \citep{wang-etal-2020-balancing} use gradient-derived rewards to adjust the training distribution online. DoGE \citep{doge} proposes a \emph{generalization estimation function} to approximate data influence, while methods like DoReMi \citep{doremi, dsdm} rely on proxy models to estimate dataset utility. In practice, these dynamic methods face two main challenges: (1) instability and high variance introduced by stochastic gradients, which we empirically demonstrate in Section~\ref{sec:results}, and (2) substantial computational overhead when proxy models or expensive estimators are required. Together, these limitations motivate the design of an online, adaptive optimization strategy that can learn sampling weights efficiently and robustly while remaining computationally tractable.

In this paper, we propose \textbf{Inf}luence-guided \textbf{D}ynamic \textbf{D}ata \textbf{S}ampling strategy \textbf{(\ouralgo{})}, a computationally efficient novel algorithm that addresses the critical challenge of data sampling for domain adaptation, overcoming limitations of existing gradient-based methods. \ouralgo{} iteratively takes small gradient-update steps on each domain’s data, monitoring the impact on the downstream metric. Domains demonstrating greater performance improvements are subsequently assigned higher rewards. This adaptive sampling strategy, inspired by recent influence-based methods \citep{pmlr-v70-koh17a, DBLP:conf/nips/BaeNLGG22, doge, mates, less} for sampling training data across multiple domains, focuses learning on the most informative subsets of data. Our algorithm offers three key advantages over prior work: (1) it eliminates the dependence on noisy gradient estimates for reward computation, (2) it efficiently reuses computations from updating the parameterized sampling distribution $\psi$ parameters to also update the model parameters $\theta$, making it much more computationally efficient and (3) it produces more reliable and interpretable sampling trajectories for better downstream gains.

Our contributions in this work are as follows:
\par\textbf{a.} We propose \ouralgo{}, an influence-guided reinforcement learning approach that learns to adjust sampling probabilities across diverse training datasets, improving target-domain retrieval performance while being much more computationally efficient.
\par\textbf{b.} We validate \ouralgo{} against robust benchmarks, including BEIR datasets \cite{DBLP:journals/corr/abs-2104-08663}, Sentence-Transformers \emph{all-MiniLM-L6-v2} \cite{reimers-gurevych-2019-sentence} and MLDR \cite{bgem3}, demonstrating significant improvements while optimizing for a target domain.

\section{Related Work}
\label{sec:related}

Recent work on training models with large, diverse data pools has explored both simple heuristic sampling and more adaptive, learned reweighting schemes. A common practice is to sample languages uniformly or via a temperature-scaled distribution that interpolates between uniform and size-proportional sampling. For example in language pretraining, Cooldown \citep{cooldown} demonstrates improvements in multilingual training by oversampling high-resource languages during the initial phases of training, while shifting to uniform sampling across high- and low-resource languages toward the end to enhance generalization. DoReMi leverages the loss gap between proxy models and the target model to optimize domain sampling weights for train set generalization.

 Early works on domain- or task-specific adaptation select relevant subsets of data using either cross-entropy differences or simple classifiers \cite{intelligent-selection, classifier-gpt3}. DSIR \citep{dsir} employs importance sampling by assigning weights to training instances based on their hashed features for the target task, which then guide data selection. In contrast, CRISP \citep{crisp} clusters the training data and assigns importance weights to clusters based on the frequency of source/target instances that fall into each cluster.

Recent influence estimation approaches guide data selection by identifying and prioritizing examples that most impact a model's predictions or performance, ensuring the most influential data are included in training \citep{influence-study, influencedistillation, davir}. Methods such as LESS \citep{less} and Quad \citep{harnessingdiversitydataselect} use first-order approximations of Influence estimates via Taylor expansions of the change in target loss after a gradient update to select relevant instances. In contrast, we focus on learning domain weights \emph{online} during training. Prior work shows that data models can accurately predict the influence of training samples on held-out target examples \citep{datamodels, dsdm}. MATES \citep{mates} similarly uses a small proxy model to locally estimate oracle influence after a single step, but small proxies may provide limited accuracy. Building on these insights, we propose leveraging online proxy models to compute exact influence scores as signals for learning sampling weights.

DDS and DoGE move beyond fixed heuristics by learning a scorer network through bi-level optimization, jointly updating the scorer and model parameters to prioritize training examples whose gradients align with a held-out development set. \citet{wang-etal-2020-balancing} extend DDS to multiple targets (MultiDDS) by learning per-language scoring functions that optimize performance across development sets. However, in practice, gradient-based strategies suffer from significant variance in the reward signal. 
In contrast, our method builds on MultiDDS and DoGE by replacing noisy gradient-based rewards with online-computed influence scores, derived from updated model parameters. This simplifies and stabilizes training by adjusting dataset-level sampling weights and reusing intermediate computations.

Extensive research has explored domain adaptation and universal generalization; providing a comprehensive review is beyond the scope of this paper. We refer interested readers to \citep{adaptive-training-bilevel-optimization, ordermatters, unimax, instructiontuningdataselection, tracin, trak, regmix, influence-study} and related works for further insights.

\section{Influence Guided Dynamic Data Sampling}
\label{sec:sys}

\begin{figure*}[ht]
  \centering
  \includegraphics[width=0.8\linewidth]{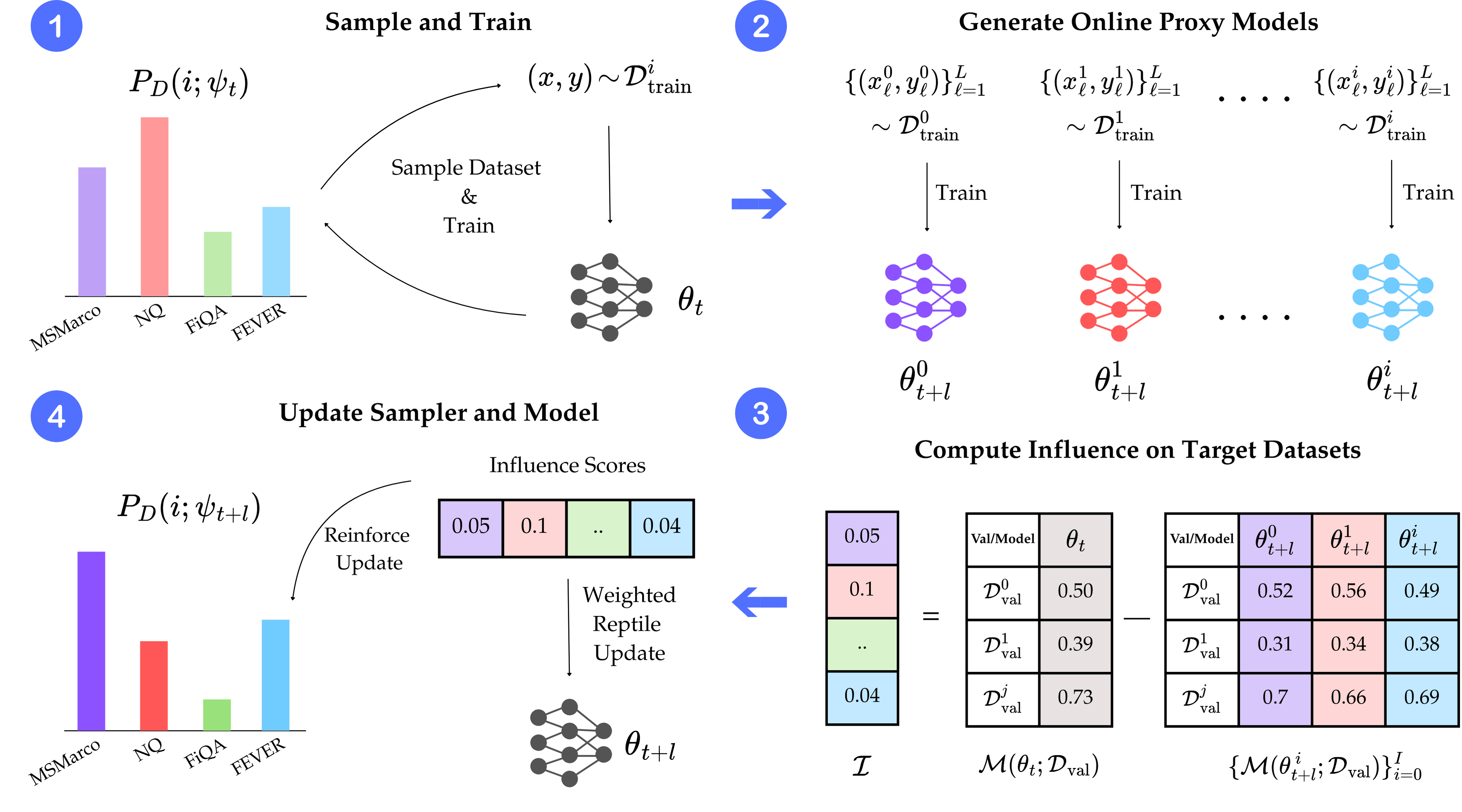}
\caption{Overview of \ouralgo{}. The trainable scorer $\psi$ and model parameters $\theta$ are optimized by generating online proxy models to compute influence scores, which serve as rewards for updating the scorer $\psi$. Proxy model gradients are efficiently reused for a weighted Reptile update on $\theta$.}
  \label{fig:mainfig}
\end{figure*}

In this section, we propose a reinforcement learning–based strategy for domain adaptation of text retrievers. We frame dataset sampling as a bilevel optimization problem and learn an adaptive policy to maximize performance on target datasets. Our approach is illustrated in Figure \ref{fig:mainfig} and elaborated in Algorithm \ref{algo:inf-dds}.

\subsection{Problem Formulation}

\textbf{Efficient data sampling for training text retrieval models:} Our goal is to devise an adaptive sampling policy that efficiently samples batches from a large pool of training domains/datasets ($M$) to maximize the model performance on a set of target datasets ($N$) a.k.a. development or dev sets.  We treat each training and dev dataset as a single homogeneous unit and optimize sampling at the dataset level.

Given a set of training datasets $\{\mathcal{D}_{\text{train}}^i\}_{i=1}^M$ with initial sampling probability as $P_D(i); i \in {1, \dots, M}$, a set of dev datasets $\{\mathcal{D}_{\text{dev}}^j\}_{j=1}^N$ on which we want our model to adapt, and initial model with parameters $\theta$, our objective is to optimize the model parameter $\theta$ by learning a dynamic sampling strategy for $P_D$ using a parameterized policy $\psi$ that maximizes performance on development sets.

We formalize our objective through the following optimization problem:
\begin{equation}
\begin{aligned}
\theta^*,\psi^* &= \argmin_{\theta,\psi} \mathbb{E}_{i \sim P_D(i;\psi)} \left[ J(\theta, \mathcal{D}_{\text{train}}^i) \right]
\end{aligned}
\label{optimization_equation}
\end{equation}
where $J(\theta, \mathcal{D}_{\text{train}}^i)$ is the empirical risk.

The training datasets sampling probability distribution $P_D(i; \psi)$ is computed as:
\begin{equation}
    P_D(i; \psi) = \frac{e^{\psi_i}}{\sum_{k=1}^M e^{\psi_k}}
\end{equation}
where $M$ is the total number of datasets in the pool. This formulation allows the sampling strategy to dynamically adjust the probabilities based on the learned parameters $(\,\psi\,)$, guiding the selection of datasets in a way that optimizes the model performance on the dev set. $(\,\psi_i\,)$ represents the importance score associated with the training dataset $i$.
\par 

We assume that the dev sets $\mathcal{D}_{\text{dev}}$ have a distribution similar to the test set $\mathcal{D}_{\text{test}}$ that is, $\mathcal{D}_{\text{dev}} \approx \mathcal{D}_{\text{test}}$. Assuming the existence of $N$ development datasets, equation \ref{optimization_equation} can be expanded as follows:
\begin{equation}
\begin{aligned}
\psi^* &= \arg\min_{\psi} \frac{1}{N} \sum_{j=1}^{N}
J\!\left(\theta^*(\psi), \mathcal{D}_{\text{dev}}^j\right) \\
\theta^* &= \arg\min_{\theta}
\mathbb{E}_{i \sim P_D(i;\psi)}
\left[ J\!\left(\theta, \mathcal{D}_{\text{train}}^i \right) \right]
\end{aligned}
\end{equation}
This formulation naturally leads to a bi-level optimization problem involving $\theta$ and $\psi$, which can be effectively addressed by alternating optimization steps. Specifically, the model parameters $\theta$ are updated using the standard gradient descent algorithm, while the scorer parameters $\psi$, are optimized using the REINFORCE algorithm \citep{DBLP:journals/ml/Williams92}.

\subsection{Influence based Rewards}

Intuitively, the reward signal should guide the parameterized scoring network $(\psi)$ to up-sample training datasets most likely to improve model performance on the development sets. We therefore propose an influence-based reward mechanism that quantifies the contribution of each training dataset to performance on a held-out development set. Existing influence-based methods \citep{less, doge, mates} either rely on first-order approximations or simulate this contribution using smaller proxy models. In contrast, we employ online proxy models to obtain accurate influence scores. Specifically, at each time step $t$, we update the model parameters $\theta_t$ using dataset $\mathcal{D}_{\text{train}}^i$ for $l$ gradient steps, where $l$ is the minimum number of steps required to reach a meaningful local minima that demonstrates the potential benefit of up-sampling $\mathcal{D}_{\text{train}}^i$.

During each of these $l$ steps, we estimate the influence of every training subset $\mathcal{D}_{\text{train}}^i$ on the current model $\theta_t$. We do this by (1) taking $l$ gradient steps on $\mathcal{D}_{\text{train}}^i$ to produce $\theta_{t+1}^i$, (2) evaluating both $\theta_t$ and $\theta_{t+1}^i$ on each dev batch using an influence metric $\mathcal{M}$, and (3) computing the change in performance:
\begin{equation}
\begin{aligned}
\Delta\mathcal{M}^i_j 
= \mathcal{M}(\theta_{t+1}^i; d_{\mathrm{val}}^j)
- \mathcal{M}(\theta_{t}^i; d_{\mathrm{val}}^j) \propto \mathcal{I}_{\theta}(i;\mathcal{D}_{\text{val}})
\end{aligned}
\end{equation}
where $\mathcal{I}_{\theta}(i;\theta)$ is the influence estimate on $\mathcal{D}_{\text{val}}$ if $\mathcal{D}_{\text{train}}^i$ is upweighed.

To ensure robustness and stability in our estimates, we normalize across the development datasets by taking a mean over all influences $\overline{\Delta\mathcal{M}^i}$ where $\overline{\Delta 
 \mathcal{M}^i} = \frac{1}{N} \sum_{j=1}^N \Delta\mathcal{M}^i_j$. We iterate the process $l$ times over all train–dev pairs $(i,j)$, accumulating the reward values $\overline{\Delta \mathcal{M}^i}$ to compute the final influence $\mathcal{I}^i$, which serves as a reliable measure of the impact of the $i^{th}$ training dataset on the model's performance on the development set.

\begin{algorithm}[ht]
\caption{Pseudocode \ouralgo{}}
\begin{algorithmic}[1]
\STATE \textbf{Input:} $\{\mathcal{D}_{\text{train}}^i\}_{i=1}^M$, $\{\mathcal{D}_{\text{val}}^j\}_{j=1}^N$, influence metric $\mathcal{M}$, inner steps per meta-update $k$
\STATE \textbf{Output:} converged model $\theta^*$
\STATE Initialize $P_D(i;\psi,\tau)\leftarrow\frac{|\mathcal{D}^i_{\text{train}}|^{1/\tau}}{\sum_{j}|\mathcal{D}^j_{\text{train}}|^{1/\tau}}$
\WHILE{$\theta_t$ not converged (every $k$ steps)}
  \STATE $\nabla_t\leftarrow0,\ S\leftarrow0$ \hfill\COMMENT{gradient cache}
  \STATE Sample val batch $\{d_{\text{val}}^j\}_{j=1}^N\sim\mathcal{D}_{\rm val}$
  \FOR{$i=1,\dots,M$} 
    \STATE  \((x,y)\sim\mathcal{D}_{\text{train}}^i\)
    \STATE $\theta_{t+1}^i\leftarrow\text{Step}(\theta_t,\mathrm{Opt}_t; x,y)$ \hfill\COMMENT{do this for $l$ steps}
    \STATE $\Delta\mathcal{M}^i_j\leftarrow\mathcal{M}(\theta_{t+1}^i;d_{\rm val}^j)-\mathcal{M}(\theta_t;d_{\rm val}^j)$ \COMMENT{compute influence}
    \STATE $\mathcal{I}^i\leftarrow\frac{1}{N}\sum_j\Delta\mathcal{M}^i_j$
    \STATE $\nabla_t \;+=\; \mathcal{I}^i(\theta_{t+1}^i-\theta_t),\quad S\;+=\;\mathcal{I}^i$ \hfill\COMMENT{accumulate influence updates}
  \ENDFOR
  \STATE $\bar\nabla_t\leftarrow\nabla_t / S$
  \STATE $\theta_{t+1}\leftarrow\theta_t+\alpha\,\bar\nabla_t$ \hfill\COMMENT{reward normalized reptile update}
  \STATE $\mathrm{Opt}_{t+1}\leftarrow\text{StateUpdate}(\mathrm{Opt}_t,\bar\nabla_t)$
  \STATE $d_\psi\leftarrow\sum_{i=1}^M P_D(i;\psi)\;\mathcal{I}^i\,\nabla_\psi\log P_D(i;\psi)$
  \STATE $\psi\leftarrow\text{GradientUpdate}(\psi,d_\psi)$ \hfill\COMMENT{sampler update}
\ENDWHILE
\end{algorithmic}
\label{algo:inf-dds}
\end{algorithm}

\subsection{Optimization for compute and scalability}
\label{subsec:opt_compute_scalability}

To reuse gradients across datasets and reduce computation, we perform Reptile-style first-order meta-updates \citep{DBLP:journals/corr/abs-1803-02999}. For each training dataset \(\mathcal{D}_{\mathrm{train}}^i\) we take \(l\) inner steps from the current initialization \(\theta_t\) (step size \(\eta_t\)), producing \(\theta_{t+1}^i\), and compute an influence score \(\mathcal{I}^i\). We convert scores to a sampling distribution via softmax, \(p_i=\exp(\mathcal{I}^i/\tau)/\sum_j\exp(\mathcal{I}^j/\tau)\), and form the weighted Reptile update
\begin{equation}
\begin{aligned}
\bar{\theta}_{t+1} &= \sum_{i=1}^M p_i\,\theta_{t+1}^i \\
\theta_{t+1} &= \theta_t + \alpha\bigl(\bar{\theta}_{t+1} - \theta_t\bigr)
\end{aligned}
\end{equation}
with Reptile rate \(\alpha=\eta_t\). Using this procedure we need only a single copy of parameter gradients and optimizer states (first/second moments), which substantially reduces memory overheads.

When \(M\) is large (e.g., many domains or languages) computing \(I^i\) for every \(i\) each iteration is costly. We therefore update the scorer \(\psi\) on a uniform random subsample \(S\subset\{\mathcal{D}_{\mathrm{train}}^i\}_{i=1}^M\) with \(|S|=k< M\). Restricting the policy to \(S\) yields the conditional categorical

\begin{equation}
\begin{aligned}
P_D(i \mid S;\psi) &=
\begin{cases}
\dfrac{P_D(i;\psi)}{\sum_{j\in S} P_D(j;\psi)}, & i \in S, \\[4pt]
0, & i \notin S ,
\end{cases} \\
P_D(i;\psi) &= \dfrac{\exp(\psi_i)}{\sum_{j=1}^{M} \exp(\psi_j)} .
\end{aligned}
\end{equation}

We then compute the scorer gradient on \(S\) as
\begin{equation}
\begin{aligned}
d_\psi \;=\; \sum_{i\in S} P_D(i;\psi)\;I^i\,\nabla_\psi\log P_D(i\mid S;\psi).
\end{aligned}
\end{equation}

This estimator is unbiased for the \emph{conditional} objective over \(S\) (by the policy-gradient identity) but biased with respect to the full-objective gradient over all \(M\) datasets. In practice, choosing \(k< M\) yields large reductions in per-iteration compute and memory while still improving performance (see Figure~\ref{fig:complexity} and Section~\ref{sec:expt}).

\section{Experimental Setup}
\label{sec:expt}

In this section, we detail the benchmarks and experimental setup used to test the domain adaptation of text retrievers under different sampling strategies.

\textbf{BEIR:}
We start with a controlled setup where in-domain train, dev, and test retrieval datasets are available, though their sizes vary across domains. We train on seven BEIR-15 datasets: MSMarco, NQ, FEVER, FiQA, HotpotQA, SciFact, and NFCorpus, first optimizing on the FEVER dev set and then extending to other datasets with available dev and test sets, including Quora, FiQA, HotpotQA, and DBpedia. NFCorpus is excluded from the target datasets due to noise, as many level-1 relevant passages are in fact irrelevant. Our biencoder models are initialized with a pretrained \emph{roberta-base} model \citep{DBLP:journals/corr/abs-1907-11692}, and trained for 2 epochs with InfoNCE loss \citep{infonce} as influence metric $\mathcal{M}$ with cross-batch negatives. The scorer network is warmed up for 50 steps and updated every 50 training steps.

\textbf{Multilingual Long Document Retrieval (MLDR):} We next examine the realistic setting of Multilingual Long Document Retrieval, using the BGE-M3 corpus originally employed to train the \emph{bge-m3-dense} model. Adaptation is performed on the MLDR-13 development set, with evaluation on its corresponding test set. The biencoder model is initialized from a 568M parameter bge-m3-unsupervised checkpoint\footnote{\href{https://huggingface.co/BAAI/bge-m3-unsupervised}{BAAI/bge-m3-unsupervised}}. We treat each language as a domain and sample proportionally across all datasets in a language. The scorer network is warmed up for 500 steps and updated every 250 training steps. We use the \emph{m3-kd-distill} loss from BGE-M3, with cross-batch negatives and 8 hard negatives, as our influence metric. 

\textbf{Sentence-Transformers Embedding Dataset:} Finally, we consider a more challenging scenario where train and test datasets span diverse domains, using the publicly available \emph{sentence-transformers embedding dataset}\footnote{\href{https://huggingface.co/collections/sentence-transformers/embedding-model-datasets-6644d7a3673a511914aa7552}{sentence-transformers/embedding-model-datasets}}, originally used to train the \emph{all-MiniLM-L6-v2} model. The training corpus contains 1 billion parallel sentences drawn from 32 datasets. Its data configuration includes carefully tuned sampling weights, referred to as \textit{Expert} initialization, designed to optimize performance. To align with our target domain, we excluded the Reddit comments dataset due to its large size and the CodeSearchNet dataset, as it is unrelated to code-focused tasks. Our biencoder models are initialized with the pretrained \emph{MiniLM-L6-H384-uncased} model, and we optimize performance jointly across all BEIR-5 dev sets. The scorer network is warmed up for 500 steps and updated every 250 steps. Following the toy setting, we use the standard InfoNCE loss as the influence metric with cross-batch negatives. Additional hyperparameters are listed in Appendix~\ref{sec:appendix-hyperparameters}.

\textbf{Baselines:} We compare our sampling algorithm against four categories of baselines: (i) static sampling methods (Temperature, Cooldown), (ii) a universal generalization approach (DoReMi), (iii) gradient-based task-adaptive sampling methods (MultiDDS, DoGE), and (iv) a cluster-level, task-adaptive importance-sampling method (CRISP). For all baselines, we follow the training guidelines recommended in their respective papers, with hyperparameter details provided in Appendix \ref{sec:appendix-hyperparameters}. We evaluate statistical significance using a paired t-test comparing the two highest-performing models in each setting. For CRISP, we construct clusters in powers of $32^x$: $x \in {1,2}$ for BEIR-train, $x \in {1,2,3}$ for Sentence-Transformers, and $x \in {1,2}$ per language for BGE-M3.

\begin{figure*}[ht]
  \centering
  \includegraphics[width=\linewidth]{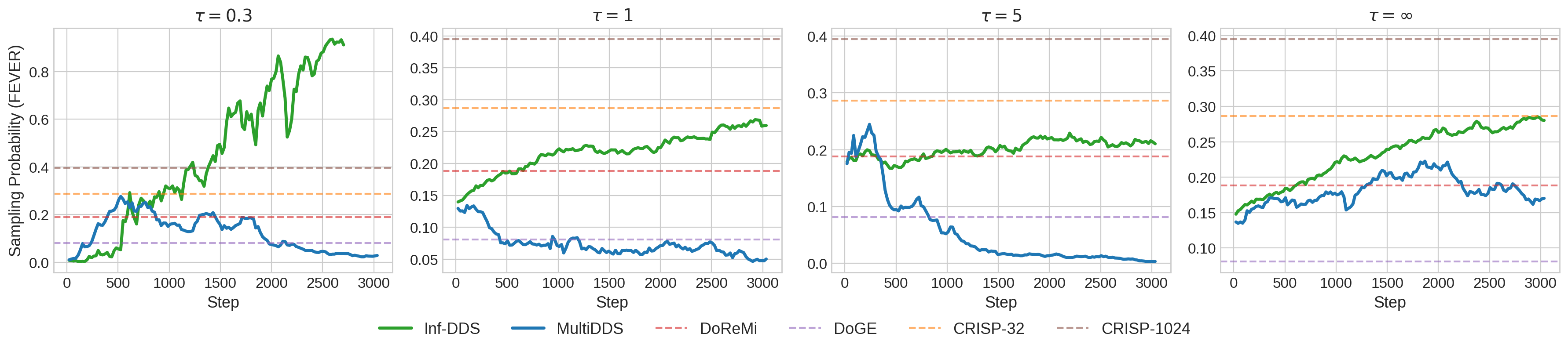}
  \caption{FEVER training set sampling trajectories for different initialization temperatures using MultiDDS, \ouralgo{}, and learned weights from baseline methods (optimized on FEVER dev set).}
  \label{fig:plotfever}
\end{figure*}

\section{Results And Analysis}
\label{sec:results}

In this section, we aim to answer the following research questions: (1) Does learning a dynamically evolving sampling distribution through influence measures lead to superior adaptation on the test set? (2) Does the influence‐based scorer capture additional insights beyond domain similarity between training and dev sets?  (3) In a diverse domain setting, how reliable are influence-based approaches compared to gradient-based methods? 

\subsection{Main Results}

\begin{figure*}[htbp]
  \centering
  \subfigure[]{%
    \label{fig:heatmapindividual}%
    \includegraphics[width=0.4\textwidth]{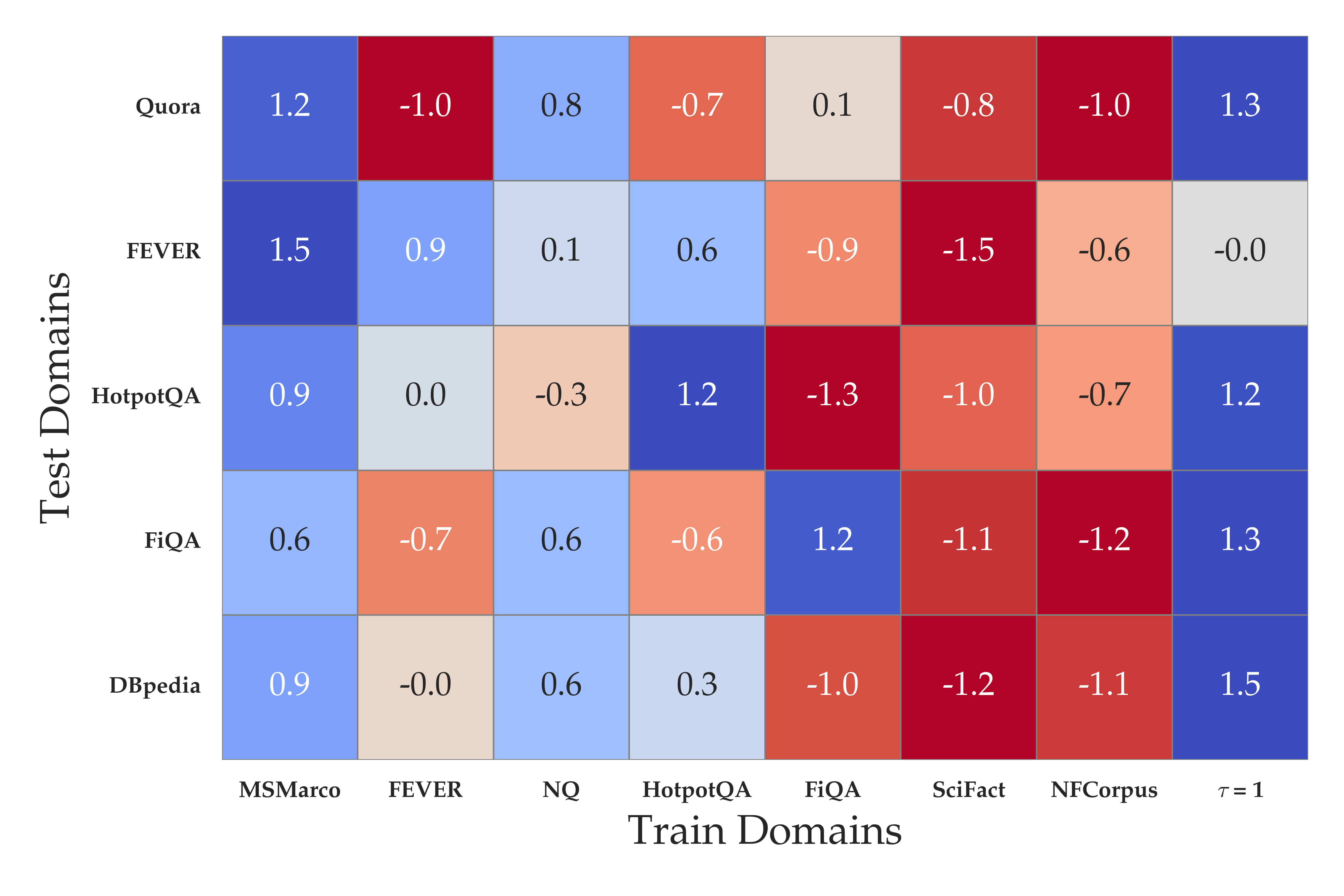}
  }
  \subfigure[]{%
    \label{fig:gridplot}%
    \includegraphics[width=0.4\textwidth]{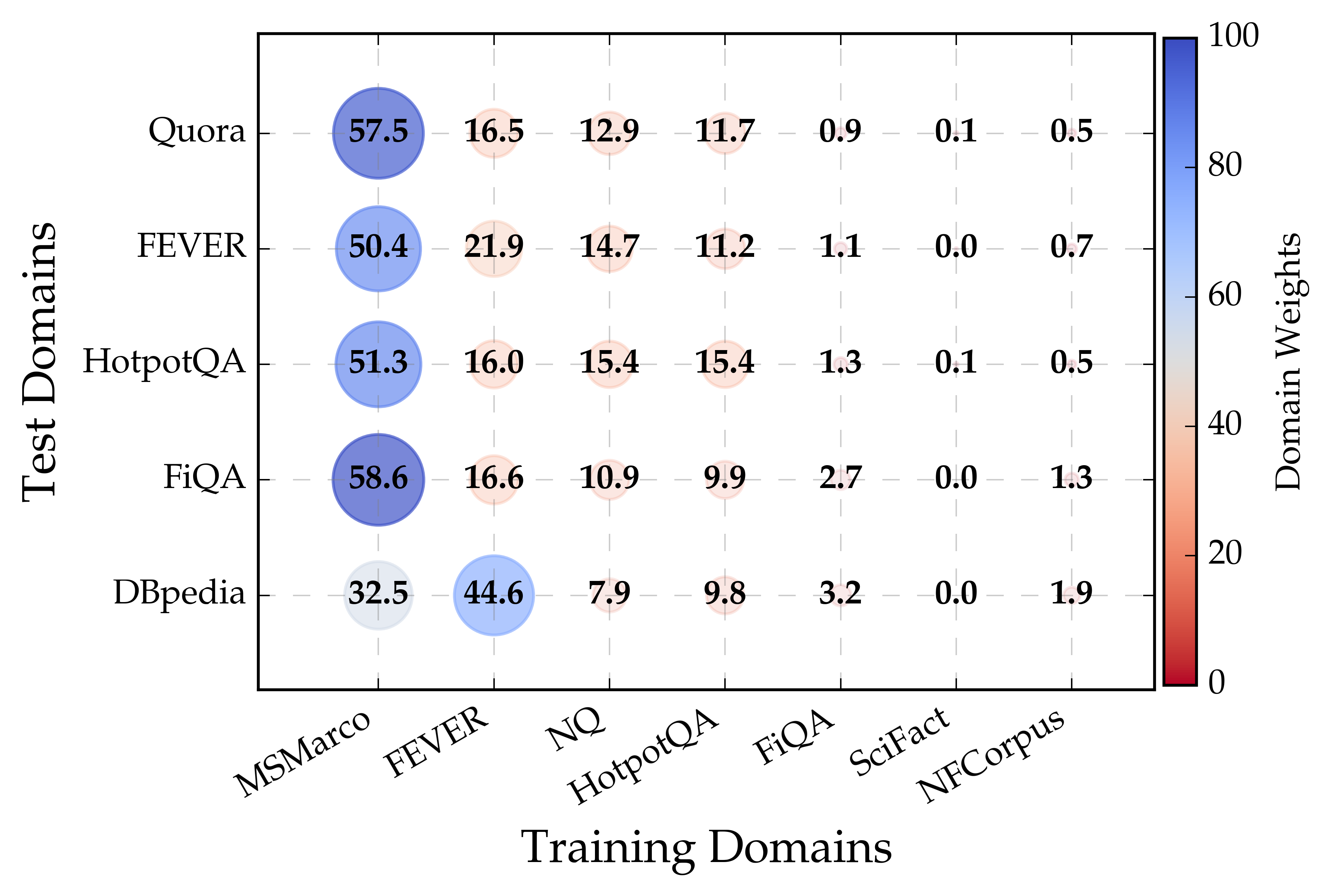}
  }
  \caption{(a) Heatmap showing Z-score (row) normalized performance correlations between train and test splits across BEIR datasets.
  (b) Domain weights learned by \ouralgo{} during optimization for each target domain.}
  \label{fig:heatmap}
\end{figure*}

\begin{figure*}[htbp]
  \centering
  \subfigure[]{%
    \label{fig:beir7trainfever}%
    \includegraphics[width=0.45\textwidth]{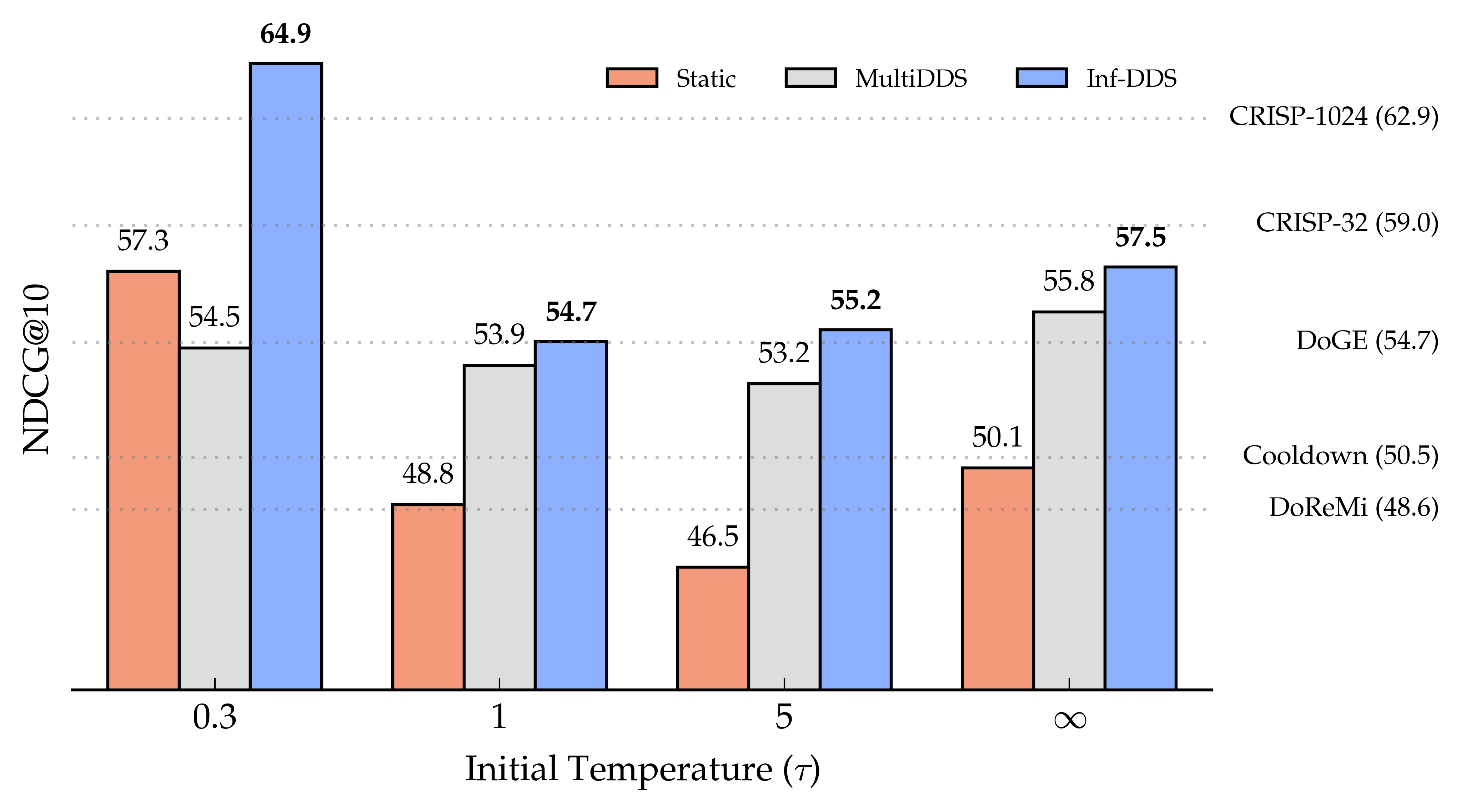}
  }
  \subfigure[]{%
    \label{fig:beir7beir5joint}%
    \includegraphics[width=0.45\textwidth]{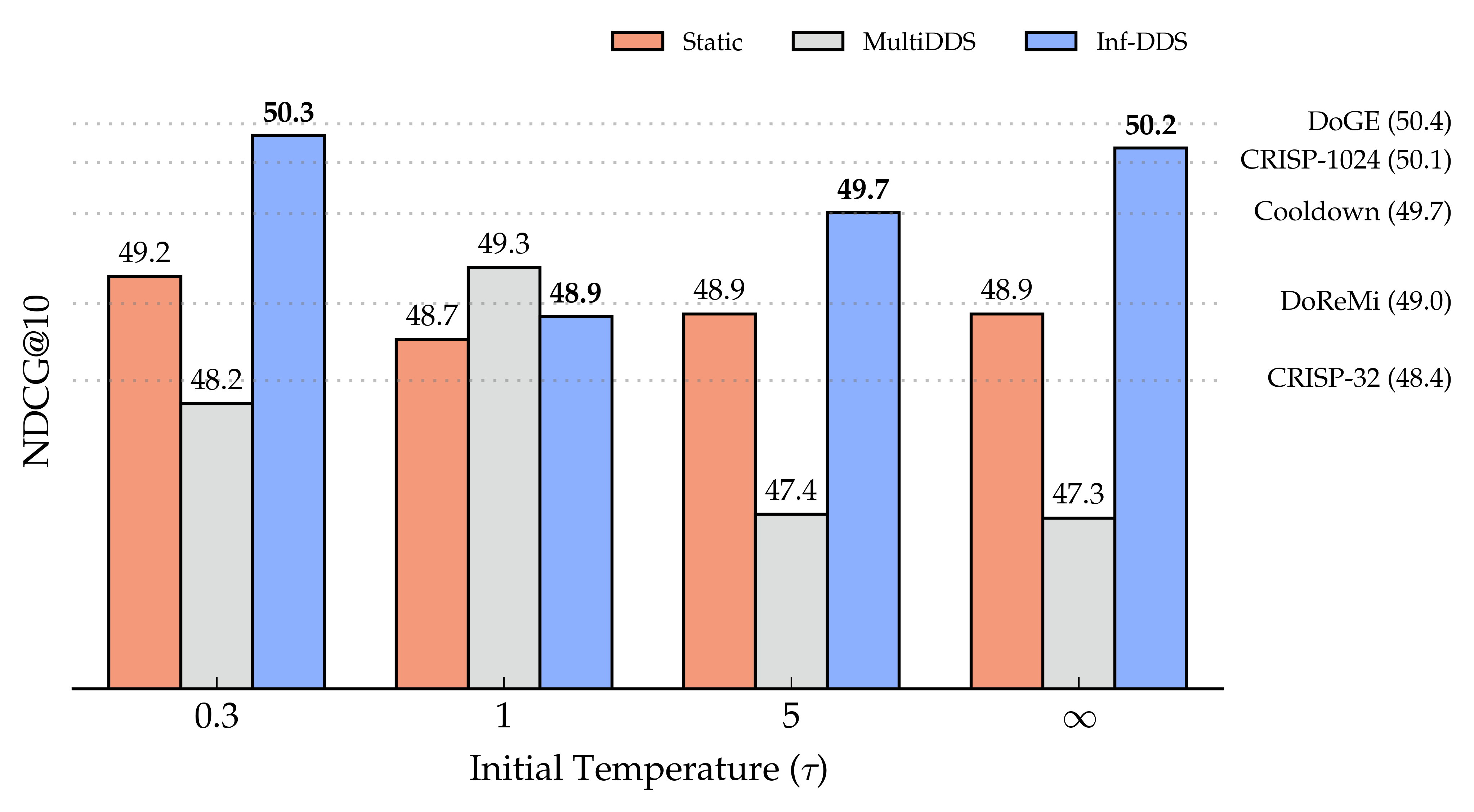}
  }
  \caption{Sampling probability initialization vs.\ Average NDCG@10 on the FEVER and BEIR-5 test set while training with BEIR-7 train set.
  (a) Scorer optimized only on the FEVER dev set.
  (b) Scorer optimized jointly on BEIR-5 dev set.}
  \label{fig:beir7}
\end{figure*}

\paragraph{Domain adaptation on \textbf{BEIR}:} Our initial findings demonstrates that mere domain similarity alone does not consistently lead to enhanced performance in text retrieval setting.
Figure~\ref{fig:heatmapindividual} illustrates this by showing the normalized performance correlation between train/test sets without adaptive sampling. Notably, target datasets such as FEVER, HotpotQA, and FiQA benefit not only from their corresponding training domains but also from MS MARCO and $\tau=1$ sampling, indicating that \textit{effective retriever improvement requires going beyond domain similarity} highlighting the need for adaptive sampling strategies designed to optimize downstream performance.

To investigate this, we adapt a sampling distribution to the FEVER development set. As \ouralgo{} relies on single-shot optimization, it is sensitive to the \textit{initial distribution}, prompting us to assess multiple initializations to ensure robustness. As shown in Figure~\ref{fig:beir7trainfever}, \ouralgo{} consistently outperforms all baselines when initialized with $\tau=0.3$, and remains competitive with CRISP using 32 clusters. Figure~\ref{fig:plotfever} provides a comparision of sampling trajectories, highlighting that MultiDDS exhibits unstable and inconsistent dynamics across varying temperatures, \ouralgo{} produces stable behavior, consistently prioritizing FEVER and MS MARCO (Figure~\ref{fig:gridplot}), in line with CRISP and DoReMi. This stability underscores the reliability of \ouralgo{} in effectively aligning sampling strategies with measurable downstream performance improvements. Full results, including statistical significance tests, are presented in \cref{tab:beir7trainfever} in the Appendix.

To further validate our approach, we conduct joint optimization over the BEIR-5 dev sets for generalization. As shown in Figure~\ref{fig:beir7beir5joint},  \ouralgo{} outperforms static sampling in all temperature initializations and surpasses MultiDDS in 2 out of 3 scenarios. While MultiDDS shows more improvements when initialized with $\tau=1$, it underperforms \ouralgo{}' best score by \(0.93\) points. \cref{tab:beir7beir5joint} in the Appendix presents the full results, including tests for statistical significance.

\paragraph{MLDR:} Multilingual retrieval introduces distinct challenges stemming from the substantial variability in language resources and heterogeneous domain distributions. We assess how \ouralgo{} and MultiDDS tackle these challenges by implicitly harmonizing data from high and low resource languages while leveraging cross-lingual relatedness without the need for explicit supervision. Specifically, we investigate two key questions: (1) Can \ouralgo{} automatically upsample underrepresented languages within a shared multilingual corpus to improve performance? (2) How critical is sampling high-resource languages when optimizing for multiple languages?

\begin{figure}[ht]
  \centering
  \includegraphics[width=0.65\linewidth]{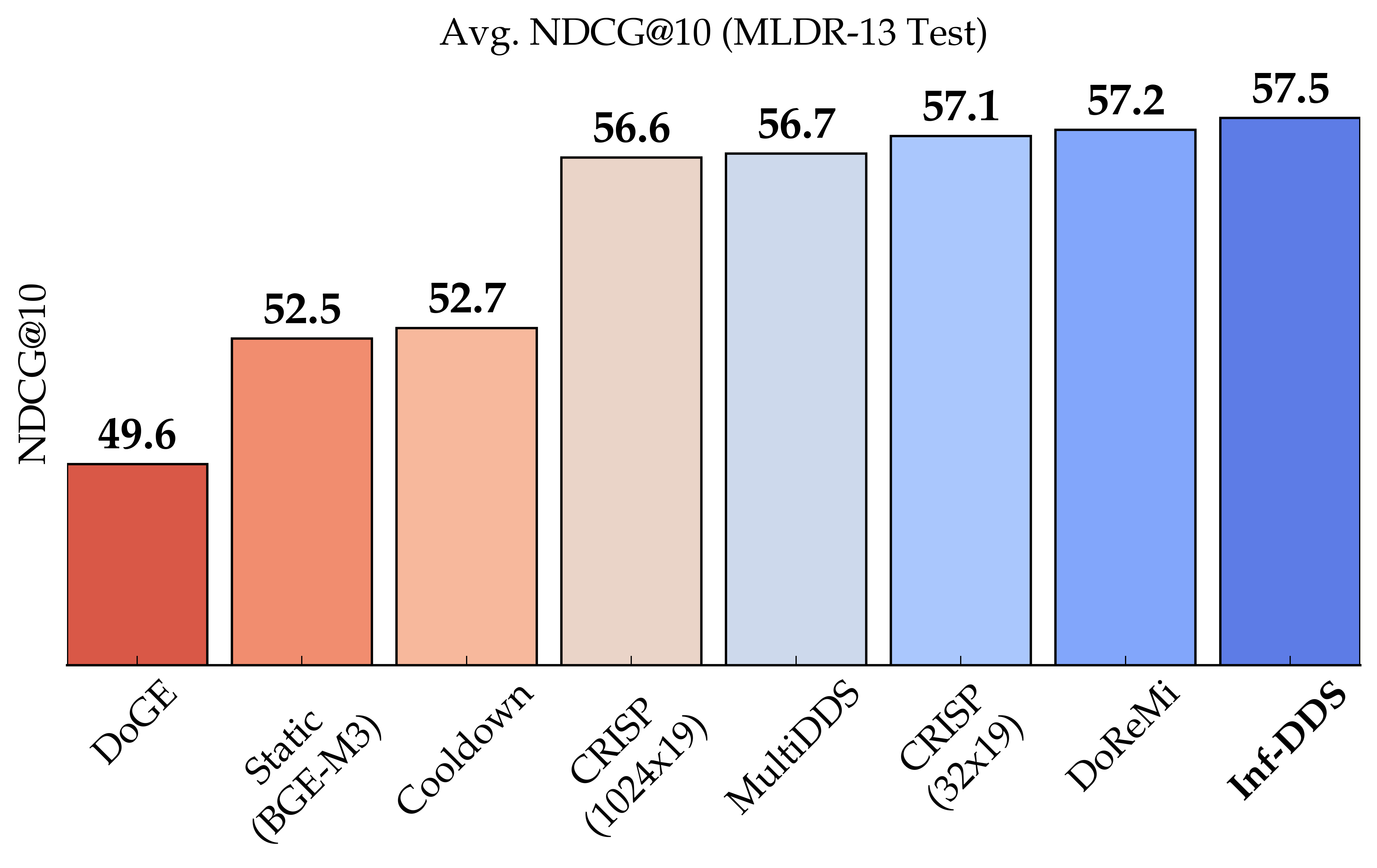}
  \caption{Avg. NDCG@10 scores on the MLDR-13 language test collection using BGE-M3 training data. Optimization of the scorer is done jointly on the 13 development sets.}
  \label{fig:mldr13}
\end{figure}

We optimize the \emph{bge-m3-unsupervised} model and scorer on the full MLDR-13 development set to maximize performance across 13 languages. As shown in Figure~\ref{fig:mldr13}, starting from the same initial sampling weights as \emph{bge-m3-dense}, \ouralgo{} improves this baseline by \(+5.03\) points in NDCG@10, while DoReMi achieves a \(+4.76\)-point gain. \ouralgo{} also achieves the highest individual-language performance in 8 out of 13 languages. Full results, including statistical significance tests, are presented in \cref{tab:mldr13} in the Appendix.

Sampling trajectories for each language are presented in Figure~\ref{fig:plotinfdds-mldr13} (Appendix). Interestingly, the sampling weights for English and Chinese drop substantially from their high initial values, yet performance remains comparable to \emph{bge-m3-dense}, likely due to their dominant presence in \emph{bge-m3-unsupervised} training (66.4\% of total). This indicates that high-resource languages need little supervised data, demonstrating dynamic sampling's ability to upweight low-resource languages while avoiding overfitting on dominant ones.

\paragraph{Sentence-Transformers Embedding Dataset:} This diverse training corpus includes over 440 Million query-positive passage pairs spanning 32 domains. We use the same BEIR-5 development sets from the toy setting for optimization, as they have minimal overlap with the training data. This experiment addresses two key questions: (1) Does dynamic sampling remain effective with high domain diversity? (2) Can our algorithm further improve performance when a strong initial sampling distribution is available? The extensive domain coverage significantly increases the complexity of the adaptation problem.

\begin{figure}[htbp]
  \centering
  \subfigure[]{%
    \includegraphics[width=0.56\linewidth]{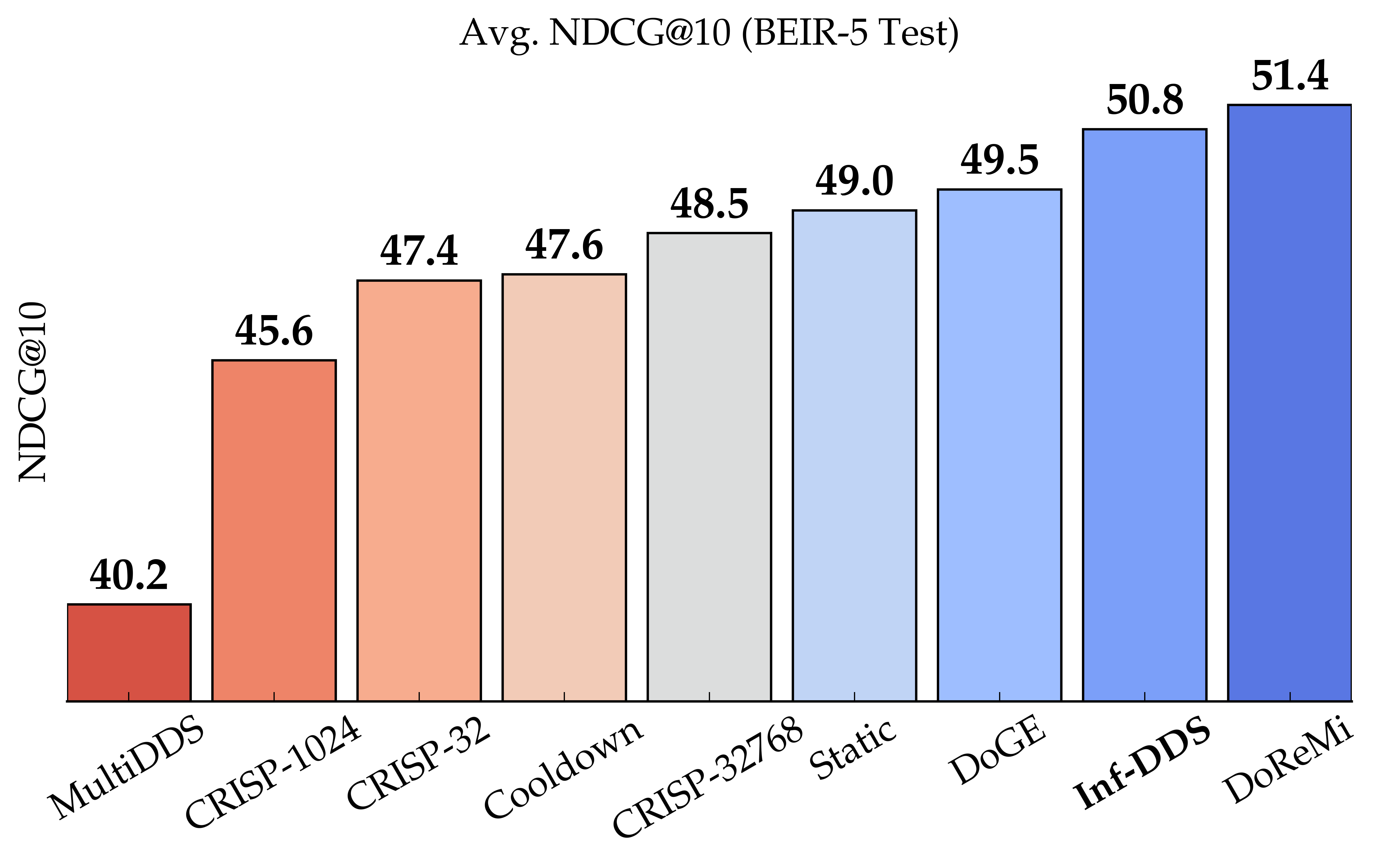}
  }
  \subfigure[]{%
    \includegraphics[width=0.38\linewidth]{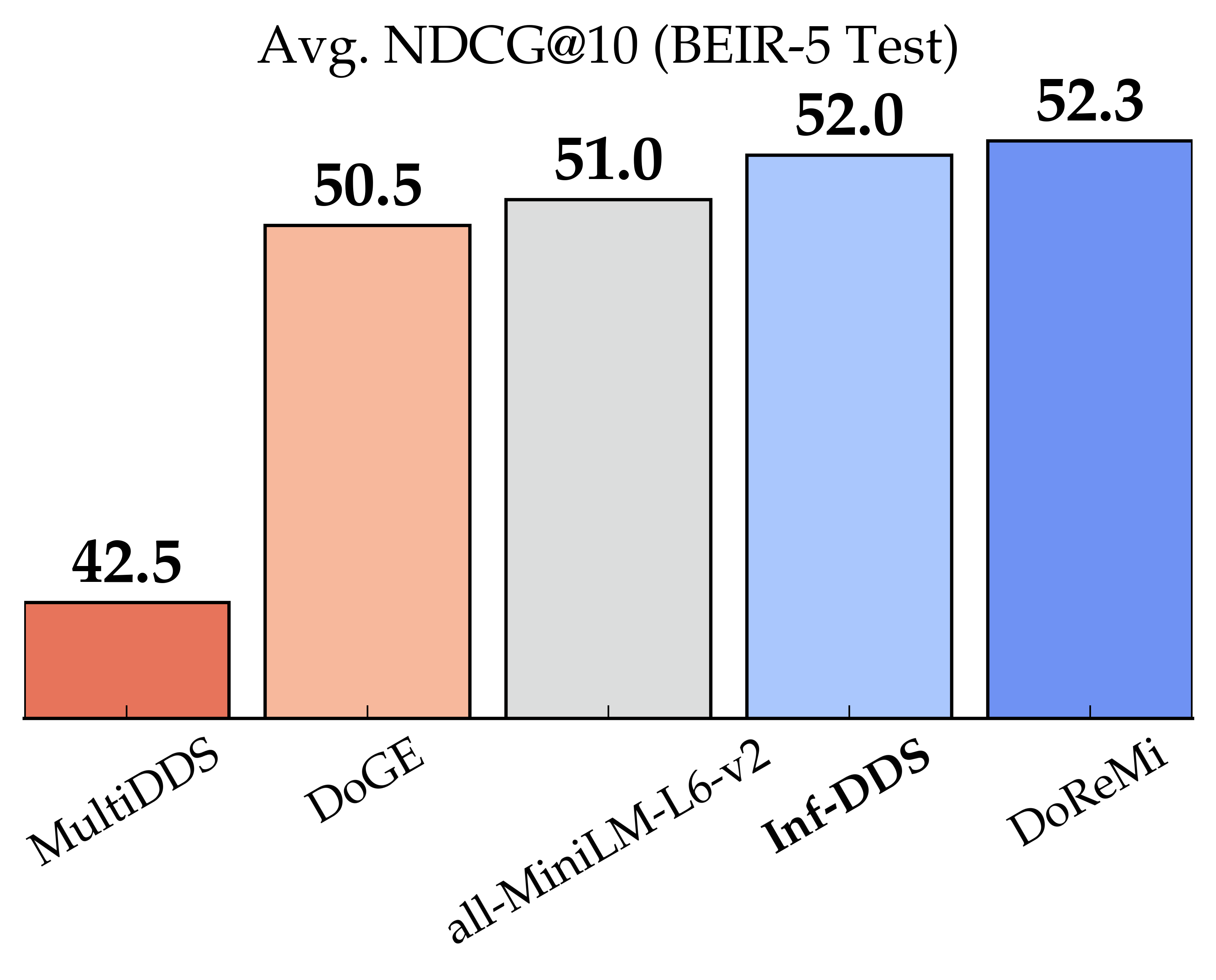}
  }
  \caption{Average NDCG@10 on the BEIR-5 test collection using Sentence-Transformers training data with uniform (a) and \textit{expert} initialization (b). The scorer is optimized jointly on the development sets.}
  \label{fig:senttransbeir5}
\end{figure}



Starting from a uniform initialization, \ouralgo{} achieves performance only \(0.22\) points below the off-the-shelf Sentence-Transformers Expert model (\textit{all-MiniLM-L6-v2}), nearly optimal—and yields a \(1.83\) point gain over the uniform baseline. In contrast, gradient-based methods such as MultiDDS and DoGE fail to make any gains. When we re-run the experiment starting from Expert weights, \ouralgo{} still produces an additional \(0.94\) point improvement, demonstrating its ability to refine even strong initial distributions. DoReMi attains a larger \(1.25\) point gain but requires 3$\times$ the compute (Fig.~\ref{fig:complexity}). Figure~\ref{fig:plotinfdds-st} shows the evolving sampling distributions under \ouralgo{}, illustrating that Expert weights can be further improved by dynamically adapting dataset sampling.

\subsection{Discussion} 

\begin{figure}[ht]
  \centering
  \includegraphics[width=0.55\linewidth]{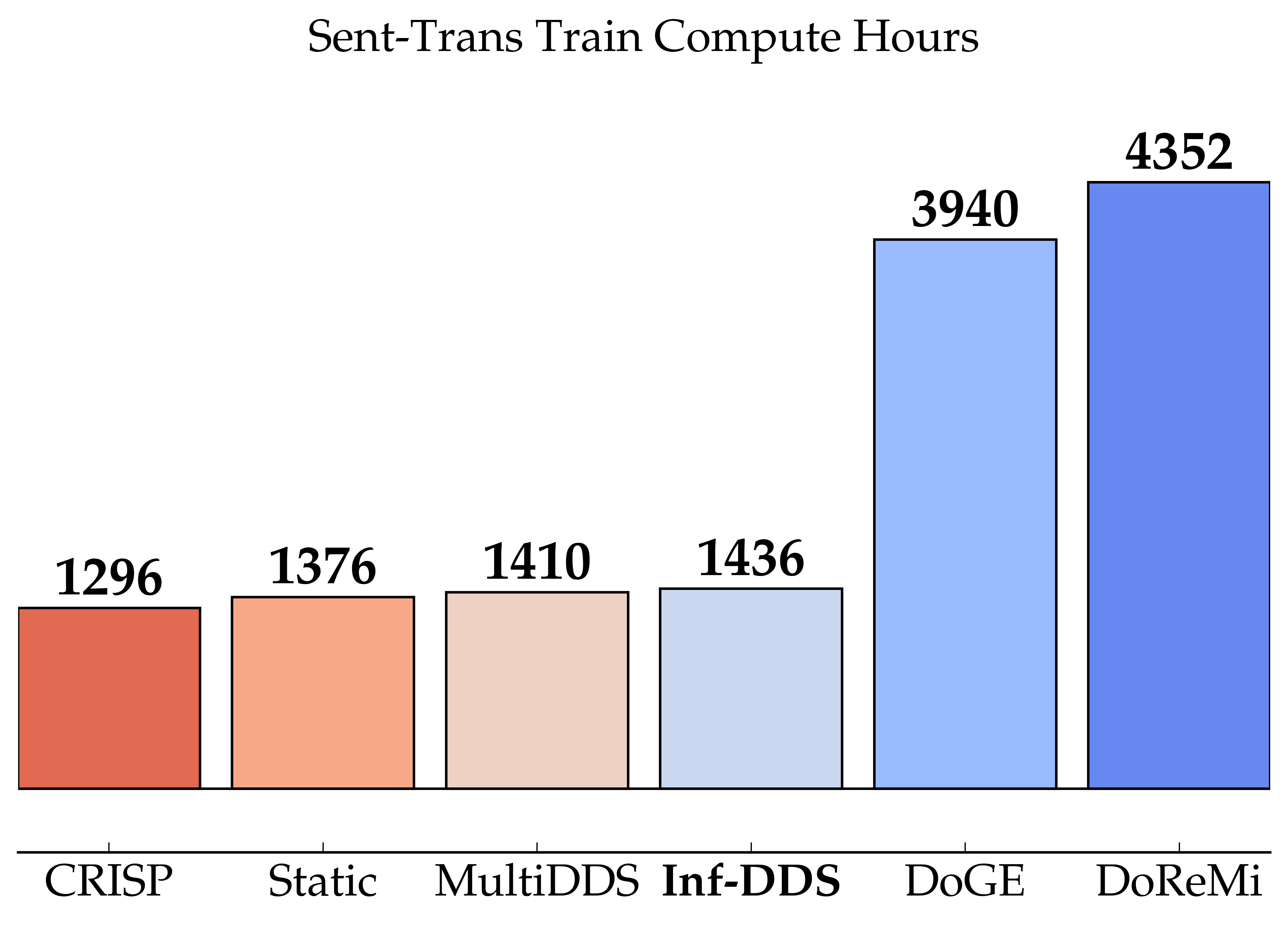}
  \caption{Comparison of approximate GPU Hours for training on Sent-Trans embedding data.}
  \label{fig:complexity}
\end{figure}

\textbf{Computational Overheads:} \ouralgo{} computes exact influence scores for each training dataset using online proxy models. While this introduces additional overhead, retrieval models are typically small, making the tradeoff worthwhile given the performance gains. Figure~\ref{fig:complexity} compares the training time of \textit{all-MiniLM-L6-v2} across different sampling strategies. Although \ouralgo{} is slightly slower than CRISP, MultiDDS, and static sampling, it consistently achieves superior performance. To reduce memory usage, \ouralgo{} stores only a single set of intermediate gradients and optimizer states during influence computation, which are efficiently reused in the weighted Reptile update.

\textbf{Effect of initialization:} Choosing an effective initialization for sampling is challenging but can substantially impact \ouralgo{}’s performance (\cref{fig:beir7trainfever,fig:beir7beir5joint,fig:senttransbeir5}). While the algorithm does not always reach globally optimal sampling weights, starting from a reasonable initialization and updating the weights consistently yields gains. We do not explore heuristics for selecting initial weights, but experiments with standard static initializations show that, although no single choice is universally best, reasonably good initializations generally perform well.

\textbf{Effect of Reptile Updates and Update Steps:} We conduct an ablation study on BEIR to evaluate the contribution of the meta-learning component, with results reported in Table~\ref{tab:ablationreptile} (Appendix). Disabling Reptile updates leads to only marginal performance changes, suggesting that the primary gains stem from dynamic sampling rather than meta-learning itself. Nevertheless, Reptile remains beneficial in practice, as it reduces computational overhead by enabling reuse of intermediate computations.
We further analyze the effect of the number of update steps ($l$) required to obtain reliable influence estimates on the FEVER and FiQA datasets under the BEIR setting. As shown in \cref{tab:update_steps}, using 3–5 update steps provides a good trade-off, yielding strong performance while avoiding unnecessary computation.

\textbf{Dev/Test Overlap Analysis:}
Several adaptive data selection and optimization methods, including DoGE, DDS, CRISP, and our approach \ouralgo{}, rely on a development set to compute rewards or influence estimates, which can raise concerns about potential dev–test leakage. While our main experiments follow the standard train, dev, and test splits provided by the original benchmarks, we also perform an analysis to verify that this protocol does not introduce test set bias.
We first train models using the original training splits of MLDR and BEIR. We then perform repeated train–dev resampling by merging training and development splits and sampling five training sets matching the original training size. All models are evaluated on the original test sets. This setup assesses whether exposure to development data during optimization affects test performance. 
Results in \cref{tab:mldr_test_leakage,tab:beir_test_leakage} show models trained on mixed train and dev folds achieve performance comparable to those trained solely on the original training data. This indicates the development sets do not leak information to the test sets and that observed gains are not driven by dev–test overlap.

\textbf{Relation between Influence and Gradients:} Stochastic gradient updates move model parameters toward the local minimum of the loss $\mathcal{L}$, whereas influence measures their impact on the target metric $\mathcal{M}$. When $\mathcal{L} \approx \mathcal{M}$, influence directly reflects the benefit of an optimization step. In contrast, gradient‑based rewards quantify the alignment between the gradient toward the dev set minimum ($\nabla_{\text{dev}}$) and the gradient from a given training instance ($\nabla_{\text{train}}$). We posit that in high-dimensional landscapes, low alignment need not indicate convergence to a poor minimum. Influence‑based rewards, by evaluating instances according to the actual target metric, provide a more direct and reliable estimate of which steps lead to the best attainable minima.

\section{Conclusion}
\label{conclusion}

In this work, we present a comprehensive analysis for adapting text retrievers to target domains using influence-guided dynamic data sampling (\ouralgo{}). Our approach parametrizes the sampling distribution with scorer parameters $\psi$ and performs bi-level optimization, jointly updating both the model parameters $\theta$ and the scorer parameters $\psi$, using influence scores as rewards. Across multiple benchmarks and baselines spanning diverse domains, \ouralgo{} produces more stable sampling trajectories and consistently comes close to or outperforms both proxy-model and gradient-based approaches, while remaining computationally efficient. Although the algorithm does not always converge to the global optimum, it reliably delivers substantial improvements from reasonable initializations. We further analyze why gradient-based signals can mislead optimization, demonstrating that influence-based rewards offer a more robust estimate of the best attainable minima. For future work, we aim to investigate improved initialization strategies and more sophisticated optimization techniques for the parameterized scorer distribution $\psi$.

\section*{Impact Statement}

This paper presents work whose goal is to advance the field of 
Machine Learning. There are many potential societal consequences 
of our work, none which we feel must be specifically highlighted here.



\begin{thebibliography}{54}
\providecommand{\natexlab}[1]{#1}
\providecommand{\url}[1]{\texttt{#1}}
\expandafter\ifx\csname urlstyle\endcsname\relax
  \providecommand{\doi}[1]{doi: #1}\else
  \providecommand{\doi}{doi: \begingroup \urlstyle{rm}\Url}\fi

\bibitem[Bae et~al.(2022)Bae, Ng, Lo, Ghassemi, and
  Grosse]{DBLP:conf/nips/BaeNLGG22}
Bae, J., Ng, N.~H., Lo, A., Ghassemi, M., and Grosse, R.~B.
\newblock If influence functions are the answer, then what is the question?
\newblock In Oh, A.~H., Agarwal, A., Belgrave, D., and Cho, K. (eds.),
  \emph{Advances in Neural Information Processing Systems}, 2022.
\newblock URL \url{https://openreview.net/forum?id=hzbguA9zMJ}.

\bibitem[BehnamGhader et~al.(2024)BehnamGhader, Adlakha, Mosbach, Bahdanau,
  Chapados, and Reddy]{DBLP:journals/corr/abs-2404-05961}
BehnamGhader, P., Adlakha, V., Mosbach, M., Bahdanau, D., Chapados, N., and
  Reddy, S.
\newblock Llm2vec: Large language models are secretly powerful text encoders.
\newblock \emph{CoRR}, abs/2404.05961, 2024.
\newblock \doi{10.48550/ARXIV.2404.05961}.
\newblock URL \url{https://doi.org/10.48550/arXiv.2404.05961}.

\bibitem[Bonifacio et~al.(2021)Bonifacio, Campiotti, Lotufo, and
  Nogueira]{DBLP:journals/corr/abs-2108-13897}
Bonifacio, L.~H., Campiotti, I., Lotufo, R.~A., and Nogueira, R.
\newblock mmarco: {A} multilingual version of {MS} {MARCO} passage ranking
  dataset.
\newblock \emph{CoRR}, abs/2108.13897, 2021.
\newblock URL \url{https://arxiv.org/abs/2108.13897}.

\bibitem[Brown et~al.(2020)Brown, Mann, Ryder, Subbiah, Kaplan, Dhariwal,
  Neelakantan, Shyam, Sastry, Askell, Agarwal, Herbert-Voss, Krueger, Henighan,
  Child, Ramesh, Ziegler, Wu, Winter, Hesse, Chen, Sigler, Litwin, Gray, Chess,
  Clark, Berner, McCandlish, Radford, Sutskever, and Amodei]{classifier-gpt3}
Brown, T.~B., Mann, B., Ryder, N., Subbiah, M., Kaplan, J., Dhariwal, P.,
  Neelakantan, A., Shyam, P., Sastry, G., Askell, A., Agarwal, S.,
  Herbert-Voss, A., Krueger, G., Henighan, T., Child, R., Ramesh, A., Ziegler,
  D.~M., Wu, J., Winter, C., Hesse, C., Chen, M., Sigler, E., Litwin, M., Gray,
  S., Chess, B., Clark, J., Berner, C., McCandlish, S., Radford, A., Sutskever,
  I., and Amodei, D.
\newblock Language models are few-shot learners.
\newblock In \emph{Proceedings of the 34th International Conference on Neural
  Information Processing Systems}, NIPS '20, Red Hook, NY, USA, 2020. Curran
  Associates Inc.
\newblock ISBN 9781713829546.

\bibitem[Cao et~al.(2024)Cao, Kang, Wang, and
  Sun]{instructiontuningdataselection}
Cao, Y., Kang, Y., Wang, C., and Sun, L.
\newblock Instruction mining: Instruction data selection for tuning large
  language models.
\newblock In \emph{First Conference on Language Modeling}, 2024.
\newblock URL \url{https://openreview.net/forum?id=wF6k0aWjAu}.

\bibitem[Chen et~al.(2024)Chen, Xiao, Zhang, Luo, Lian, and Liu]{bgem3}
Chen, J., Xiao, S., Zhang, P., Luo, K., Lian, D., and Liu, Z.
\newblock {BGE} m3-embedding: Multi-lingual, multi-functionality,
  multi-granularity text embeddings through self-knowledge distillation.
\newblock \emph{CoRR}, abs/2402.03216, 2024.
\newblock \doi{10.48550/ARXIV.2402.03216}.
\newblock URL \url{https://doi.org/10.48550/arXiv.2402.03216}.

\bibitem[Choi et~al.(2023)Choi, Xin, Dadkhahi, Gilmer, Garg, Firat, Yeh, Dai,
  and Ghorbani]{ordermatters}
Choi, D., Xin, D., Dadkhahi, H., Gilmer, J., Garg, A., Firat, O., Yeh, C.-K.,
  Dai, A.~M., and Ghorbani, B.
\newblock Order matters in the presence of dataset imbalance for multilingual
  learning.
\newblock In \emph{Thirty-seventh Conference on Neural Information Processing
  Systems}, 2023.
\newblock URL \url{https://openreview.net/forum?id=7RMGI4slcb}.

\bibitem[Chung et~al.(2023)Chung, Garcia, Roberts, Tay, Firat, Narang, and
  Constant]{unimax}
Chung, H.~W., Garcia, X., Roberts, A., Tay, Y., Firat, O., Narang, S., and
  Constant, N.
\newblock Unimax: Fairer and more effective language sampling for large-scale
  multilingual pretraining.
\newblock In \emph{The Eleventh International Conference on Learning
  Representations, {ICLR} 2023, Kigali, Rwanda, May 1-5, 2023}. OpenReview.net,
  2023.
\newblock URL \url{https://openreview.net/forum?id=kXwdL1cWOAi}.

\bibitem[Engstrom et~al.(2024)Engstrom, Feldmann, and Madry]{dsdm}
Engstrom, L., Feldmann, A., and Madry, A.
\newblock Dsdm: model-aware dataset selection with datamodels.
\newblock In \emph{Proceedings of the 41st International Conference on Machine
  Learning}, ICML'24. JMLR.org, 2024.

\bibitem[Fan et~al.(2024)Fan, Pagliardini, and Jaggi]{doge}
Fan, S., Pagliardini, M., and Jaggi, M.
\newblock Doge: domain reweighting with generalization estimation.
\newblock In \emph{Proceedings of the 41st International Conference on Machine
  Learning}, ICML'24. JMLR.org, 2024.

\bibitem[Gao et~al.(2021)Gao, Yao, and Chen]{gao-etal-2021-simcse}
Gao, T., Yao, X., and Chen, D.
\newblock {S}im{CSE}: Simple contrastive learning of sentence embeddings.
\newblock In Moens, M.-F., Huang, X., Specia, L., and Yih, S. W.-t. (eds.),
  \emph{Proceedings of the 2021 Conference on Empirical Methods in Natural
  Language Processing}, pp.\  6894--6910, Online and Punta Cana, Dominican
  Republic, November 2021. Association for Computational Linguistics.
\newblock \doi{10.18653/v1/2021.emnlp-main.552}.
\newblock URL \url{https://aclanthology.org/2021.emnlp-main.552/}.

\bibitem[Grangier et~al.(2025{\natexlab{a}})Grangier, Ablin, and
  Hannun]{adaptive-training-bilevel-optimization}
Grangier, D., Ablin, P., and Hannun, A.
\newblock Adaptive training distributions with scalable online bilevel
  optimization.
\newblock In \emph{Transactions on Machine Learning Research (TMLR)},
  2025{\natexlab{a}}.
\newblock URL \url{https://arxiv.org/abs/2311.11973}.

\bibitem[Grangier et~al.(2025{\natexlab{b}})Grangier, Fan, Seto, and
  Ablin]{crisp}
Grangier, D., Fan, S., Seto, S., and Ablin, P.
\newblock Task-adaptive pretrained language models via clustered-importance
  sampling.
\newblock In \emph{The Thirteenth International Conference on Learning
  Representations}, 2025{\natexlab{b}}.
\newblock URL \url{https://openreview.net/forum?id=p6ncr0eTKE}.

\bibitem[Grosse et~al.(2023)Grosse, Bae, Anil, Elhage, Tamkin, Tajdini,
  Steiner, Li, Durmus, Perez, Hubinger, Lukosiute, Nguyen, Joseph, McCandlish,
  Kaplan, and Bowman]{influence-study}
Grosse, R.~B., Bae, J., Anil, C., Elhage, N., Tamkin, A., Tajdini, A., Steiner,
  B., Li, D., Durmus, E., Perez, E., Hubinger, E., Lukosiute, K., Nguyen, K.,
  Joseph, N., McCandlish, S., Kaplan, J., and Bowman, S.~R.
\newblock Studying large language model generalization with influence
  functions.
\newblock \emph{CoRR}, abs/2308.03296, 2023.
\newblock \doi{10.48550/ARXIV.2308.03296}.
\newblock URL \url{https://doi.org/10.48550/arXiv.2308.03296}.

\bibitem[He et~al.(2018)He, Liu, Liu, Lyu, Zhao, Xiao, Liu, Wang, Wu, She, Liu,
  Wu, and Wang]{he-etal-2018-dureader}
He, W., Liu, K., Liu, J., Lyu, Y., Zhao, S., Xiao, X., Liu, Y., Wang, Y., Wu,
  H., She, Q., Liu, X., Wu, T., and Wang, H.
\newblock {D}u{R}eader: a {C}hinese machine reading comprehension dataset from
  real-world applications.
\newblock In Choi, E., Seo, M., Chen, D., Jia, R., and Berant, J. (eds.),
  \emph{Proceedings of the Workshop on Machine Reading for Question Answering},
  pp.\  37--46, Melbourne, Australia, July 2018. Association for Computational
  Linguistics.
\newblock \doi{10.18653/v1/W18-2605}.
\newblock URL \url{https://aclanthology.org/W18-2605/}.

\bibitem[Ilyas et~al.(2022)Ilyas, Park, Engstrom, Leclerc, and
  Madry]{datamodels}
Ilyas, A., Park, S.~M., Engstrom, L., Leclerc, G., and Madry, A.
\newblock Datamodels: Predicting predictions from training data.
\newblock \emph{CoRR}, abs/2202.00622, 2022.
\newblock URL \url{https://arxiv.org/abs/2202.00622}.

\bibitem[Izacard et~al.(2022)Izacard, Caron, Hosseini, Riedel, Bojanowski,
  Joulin, and Grave]{DBLP:journals/tmlr/IzacardCHRBJG22}
Izacard, G., Caron, M., Hosseini, L., Riedel, S., Bojanowski, P., Joulin, A.,
  and Grave, E.
\newblock Unsupervised dense information retrieval with contrastive learning.
\newblock \emph{Trans. Mach. Learn. Res.}, 2022, 2022.
\newblock URL \url{https://openreview.net/forum?id=jKN1pXi7b0}.

\bibitem[Jin et~al.(2019)Jin, Dhingra, Liu, Cohen, and
  Lu]{jin-etal-2019-pubmedqa}
Jin, Q., Dhingra, B., Liu, Z., Cohen, W., and Lu, X.
\newblock {P}ub{M}ed{QA}: A dataset for biomedical research question answering.
\newblock In Inui, K., Jiang, J., Ng, V., and Wan, X. (eds.), \emph{Proceedings
  of the 2019 Conference on Empirical Methods in Natural Language Processing
  and the 9th International Joint Conference on Natural Language Processing
  (EMNLP-IJCNLP)}, pp.\  2567--2577, Hong Kong, China, November 2019.
  Association for Computational Linguistics.
\newblock \doi{10.18653/v1/D19-1259}.
\newblock URL \url{https://aclanthology.org/D19-1259/}.

\bibitem[Joshi et~al.(2017)Joshi, Choi, Weld, and
  Zettlemoyer]{joshi-etal-2017-triviaqa}
Joshi, M., Choi, E., Weld, D., and Zettlemoyer, L.
\newblock {T}rivia{QA}: A large scale distantly supervised challenge dataset
  for reading comprehension.
\newblock In Barzilay, R. and Kan, M.-Y. (eds.), \emph{Proceedings of the 55th
  Annual Meeting of the Association for Computational Linguistics (Volume 1:
  Long Papers)}, pp.\  1601--1611, Vancouver, Canada, July 2017. Association
  for Computational Linguistics.
\newblock \doi{10.18653/v1/P17-1147}.
\newblock URL \url{https://aclanthology.org/P17-1147/}.

\bibitem[Karpukhin et~al.(2020)Karpukhin, Oguz, Min, Lewis, Wu, Edunov, Chen,
  and Yih]{karpukhin-etal-2020-dense}
Karpukhin, V., Oguz, B., Min, S., Lewis, P., Wu, L., Edunov, S., Chen, D., and
  Yih, W.-t.
\newblock Dense passage retrieval for open-domain question answering.
\newblock In Webber, B., Cohn, T., He, Y., and Liu, Y. (eds.),
  \emph{Proceedings of the 2020 Conference on Empirical Methods in Natural
  Language Processing (EMNLP)}, pp.\  6769--6781, Online, November 2020.
  Association for Computational Linguistics.
\newblock \doi{10.18653/v1/2020.emnlp-main.550}.
\newblock URL \url{https://aclanthology.org/2020.emnlp-main.550/}.

\bibitem[Kim et~al.(2022)Kim, Rabelo, Goebel, Yoshioka, Kano, and
  Satoh]{10.1007/978-3-031-29168-5_4}
Kim, M.-Y., Rabelo, J., Goebel, R., Yoshioka, M., Kano, Y., and Satoh, K.
\newblock Coliee 2022 summary: Methods for legal document retrieval and
  entailment.
\newblock In \emph{New Frontiers in Artificial Intelligence: JSAI-IsAI 2022
  Workshop, JURISIN 2022, and JSAI 2022 International Session, Kyoto, Japan,
  June 12–17, 2022, Revised Selected Papers}, pp.\  51–67, Berlin,
  Heidelberg, 2022. Springer-Verlag.
\newblock ISBN 978-3-031-29167-8.
\newblock \doi{10.1007/978-3-031-29168-5_4}.
\newblock URL \url{https://doi.org/10.1007/978-3-031-29168-5_4}.

\bibitem[Koh \& Liang(2017)Koh and Liang]{pmlr-v70-koh17a}
Koh, P.~W. and Liang, P.
\newblock Understanding black-box predictions via influence functions.
\newblock In Precup, D. and Teh, Y.~W. (eds.), \emph{Proceedings of the 34th
  International Conference on Machine Learning}, volume~70 of \emph{Proceedings
  of Machine Learning Research}, pp.\  1885--1894. PMLR, 06--11 Aug 2017.
\newblock URL \url{https://proceedings.mlr.press/v70/koh17a.html}.

\bibitem[Kwiatkowski et~al.(2019)Kwiatkowski, Palomaki, Redfield, Collins,
  Parikh, Alberti, Epstein, Polosukhin, Devlin, Lee, Toutanova, Jones, Kelcey,
  Chang, Dai, Uszkoreit, Le, and Petrov]{kwiatkowski-etal-2019-natural}
Kwiatkowski, T., Palomaki, J., Redfield, O., Collins, M., Parikh, A., Alberti,
  C., Epstein, D., Polosukhin, I., Devlin, J., Lee, K., Toutanova, K., Jones,
  L., Kelcey, M., Chang, M.-W., Dai, A.~M., Uszkoreit, J., Le, Q., and Petrov,
  S.
\newblock Natural questions: A benchmark for question answering research.
\newblock \emph{Transactions of the Association for Computational Linguistics},
  7:\penalty0 452--466, 2019.
\newblock \doi{10.1162/tacl_a_00276}.
\newblock URL \url{https://aclanthology.org/Q19-1026/}.

\bibitem[Langley(2000)]{langley00}
Langley, P.
\newblock Crafting papers on machine learning.
\newblock In Langley, P. (ed.), \emph{Proceedings of the 17th International
  Conference on Machine Learning (ICML 2000)}, pp.\  1207--1216, Stanford, CA,
  2000. Morgan Kaufmann.

\bibitem[Li et~al.(2024{\natexlab{a}})Li, Shao, Wu, Ai, Ma, and
  Liu]{10.1145/3626772.3657887}
Li, H., Shao, Y., Wu, Y., Ai, Q., Ma, Y., and Liu, Y.
\newblock Lecardv2: A large-scale chinese legal case retrieval dataset.
\newblock In \emph{Proceedings of the 47th International ACM SIGIR Conference
  on Research and Development in Information Retrieval}, SIGIR '24, pp.\
  2251–2260, New York, NY, USA, 2024{\natexlab{a}}. Association for Computing
  Machinery.
\newblock ISBN 9798400704314.
\newblock \doi{10.1145/3626772.3657887}.
\newblock URL \url{https://doi.org/10.1145/3626772.3657887}.

\bibitem[Li et~al.(2024{\natexlab{b}})Li, Xu, Tan, Murray, and
  Khashabi]{cooldown}
Li, T., Xu, H., Tan, W., Murray, K., and Khashabi, D.
\newblock Upsample or upweight? balanced training on heavily imbalanced
  datasets.
\newblock \emph{CoRR}, abs/2410.04579, 2024{\natexlab{b}}.
\newblock \doi{10.48550/ARXIV.2410.04579}.
\newblock URL \url{https://doi.org/10.48550/arXiv.2410.04579}.

\bibitem[Liu et~al.(2025)Liu, Zheng, Muennighoff, Zeng, Dou, Pang, Jiang, and
  Lin]{regmix}
Liu, Q., Zheng, X., Muennighoff, N., Zeng, G., Dou, L., Pang, T., Jiang, J.,
  and Lin, M.
\newblock Regmix: Data mixture as regression for language model pre-training.
\newblock In \emph{The Thirteenth International Conference on Learning
  Representations}, 2025.
\newblock URL \url{https://openreview.net/forum?id=5BjQOUXq7i}.

\bibitem[Liu et~al.(2019)Liu, Ott, Goyal, Du, Joshi, Chen, Levy, Lewis,
  Zettlemoyer, and Stoyanov]{DBLP:journals/corr/abs-1907-11692}
Liu, Y., Ott, M., Goyal, N., Du, J., Joshi, M., Chen, D., Levy, O., Lewis, M.,
  Zettlemoyer, L., and Stoyanov, V.
\newblock Roberta: {A} robustly optimized {BERT} pretraining approach.
\newblock \emph{CoRR}, abs/1907.11692, 2019.
\newblock URL \url{http://arxiv.org/abs/1907.11692}.

\bibitem[Moore \& Lewis(2010)Moore and Lewis]{intelligent-selection}
Moore, R.~C. and Lewis, W.
\newblock Intelligent selection of language model training data.
\newblock In Haji{\v{c}}, J., Carberry, S., Clark, S., and Nivre, J. (eds.),
  \emph{Proceedings of the {ACL} 2010 Conference Short Papers}, pp.\  220--224,
  Uppsala, Sweden, July 2010. Association for Computational Linguistics.
\newblock URL \url{https://aclanthology.org/P10-2041/}.

\bibitem[Nguyen et~al.(2016)Nguyen, Rosenberg, Song, Gao, Tiwary, Majumder, and
  Deng]{msmarco}
Nguyen, T., Rosenberg, M., Song, X., Gao, J., Tiwary, S., Majumder, R., and
  Deng, L.
\newblock {MS} {MARCO:} {A} human generated machine reading comprehension
  dataset.
\newblock In Besold, T.~R., Bordes, A., d'Avila Garcez, A.~S., and Wayne, G.
  (eds.), \emph{Proceedings of the Workshop on Cognitive Computation:
  Integrating neural and symbolic approaches 2016 co-located with the 30th
  Annual Conference on Neural Information Processing Systems {(NIPS} 2016),
  Barcelona, Spain, December 9, 2016}, volume 1773 of \emph{{CEUR} Workshop
  Proceedings}. CEUR-WS.org, 2016.
\newblock URL \url{https://ceur-ws.org/Vol-1773/CoCoNIPS\_2016\_paper9.pdf}.

\bibitem[Nichol et~al.(2018)Nichol, Achiam, and
  Schulman]{DBLP:journals/corr/abs-1803-02999}
Nichol, A., Achiam, J., and Schulman, J.
\newblock On first-order meta-learning algorithms.
\newblock \emph{CoRR}, abs/1803.02999, 2018.
\newblock URL \url{http://arxiv.org/abs/1803.02999}.

\bibitem[Nikdan et~al.(2025)Nikdan, Cohen-Addad, Alistarh, and
  Mirrokni]{influencedistillation}
Nikdan, M., Cohen-Addad, V., Alistarh, D., and Mirrokni, V.
\newblock Efficient data selection at scale via influence distillation.
\newblock \emph{arXiv preprint arXiv:2505.19051}, 2025.

\bibitem[Park et~al.(2023)Park, Georgiev, Ilyas, Leclerc, and Madry]{trak}
Park, S.~M., Georgiev, K., Ilyas, A., Leclerc, G., and Madry, A.
\newblock Trak: attributing model behavior at scale.
\newblock In \emph{Proceedings of the 40th International Conference on Machine
  Learning}, ICML'23. JMLR.org, 2023.

\bibitem[Pruthi et~al.(2020)Pruthi, Liu, Kale, and Sundararajan]{tracin}
Pruthi, G., Liu, F., Kale, S., and Sundararajan, M.
\newblock Estimating training data influence by tracing gradient descent.
\newblock In Larochelle, H., Ranzato, M., Hadsell, R., Balcan, M., and Lin, H.
  (eds.), \emph{Advances in Neural Information Processing Systems 33: Annual
  Conference on Neural Information Processing Systems 2020, NeurIPS 2020,
  December 6-12, 2020, virtual}, 2020.
\newblock URL
  \url{https://proceedings.neurips.cc/paper/2020/hash/e6385d39ec9394f2f3a354d9d2b88eec-Abstract.html}.

\bibitem[Rajpurkar et~al.(2016)Rajpurkar, Zhang, Lopyrev, and
  Liang]{rajpurkar-etal-2016-squad}
Rajpurkar, P., Zhang, J., Lopyrev, K., and Liang, P.
\newblock {SQ}u{AD}: 100,000+ questions for machine comprehension of text.
\newblock In Su, J., Duh, K., and Carreras, X. (eds.), \emph{Proceedings of the
  2016 Conference on Empirical Methods in Natural Language Processing}, pp.\
  2383--2392, Austin, Texas, November 2016. Association for Computational
  Linguistics.
\newblock \doi{10.18653/v1/D16-1264}.
\newblock URL \url{https://aclanthology.org/D16-1264/}.

\bibitem[Reimers \& Gurevych(2019)Reimers and
  Gurevych]{reimers-gurevych-2019-sentence}
Reimers, N. and Gurevych, I.
\newblock Sentence-{BERT}: Sentence embeddings using {S}iamese {BERT}-networks.
\newblock In Inui, K., Jiang, J., Ng, V., and Wan, X. (eds.), \emph{Proceedings
  of the 2019 Conference on Empirical Methods in Natural Language Processing
  and the 9th International Joint Conference on Natural Language Processing
  (EMNLP-IJCNLP)}, pp.\  3982--3992, Hong Kong, China, November 2019.
  Association for Computational Linguistics.
\newblock \doi{10.18653/v1/D19-1410}.
\newblock URL \url{https://aclanthology.org/D19-1410/}.

\bibitem[Thakur et~al.(2021)Thakur, Reimers, R{\"{u}}ckl{\'{e}}, Srivastava,
  and Gurevych]{DBLP:journals/corr/abs-2104-08663}
Thakur, N., Reimers, N., R{\"{u}}ckl{\'{e}}, A., Srivastava, A., and Gurevych,
  I.
\newblock {BEIR:} {A} heterogenous benchmark for zero-shot evaluation of
  information retrieval models.
\newblock \emph{CoRR}, abs/2104.08663, 2021.
\newblock URL \url{https://arxiv.org/abs/2104.08663}.

\bibitem[van~den Oord et~al.(2018)van~den Oord, Li, and Vinyals]{infonce}
van~den Oord, A., Li, Y., and Vinyals, O.
\newblock Representation learning with contrastive predictive coding.
\newblock \emph{CoRR}, abs/1807.03748, 2018.
\newblock URL \url{http://arxiv.org/abs/1807.03748}.

\bibitem[Wang et~al.(2023)Wang, Yang, Huang, Jiao, Yang, Jiang, Majumder, and
  Wei]{wang-etal-2023-simlm}
Wang, L., Yang, N., Huang, X., Jiao, B., Yang, L., Jiang, D., Majumder, R., and
  Wei, F.
\newblock {S}im{LM}: Pre-training with representation bottleneck for dense
  passage retrieval.
\newblock In Rogers, A., Boyd-Graber, J., and Okazaki, N. (eds.),
  \emph{Proceedings of the 61st Annual Meeting of the Association for
  Computational Linguistics (Volume 1: Long Papers)}, pp.\  2244--2258,
  Toronto, Canada, July 2023. Association for Computational Linguistics.
\newblock \doi{10.18653/v1/2023.acl-long.125}.
\newblock URL \url{https://aclanthology.org/2023.acl-long.125/}.

\bibitem[Wang et~al.(2020{\natexlab{a}})Wang, Pham, Michel, Anastasopoulos,
  Carbonell, and Neubig]{DDS}
Wang, X., Pham, H., Michel, P., Anastasopoulos, A., Carbonell, J.~G., and
  Neubig, G.
\newblock Optimizing data usage via differentiable rewards.
\newblock In \emph{Proceedings of the 37th International Conference on Machine
  Learning, {ICML} 2020, 13-18 July 2020, Virtual Event}, volume 119 of
  \emph{Proceedings of Machine Learning Research}, pp.\  9983--9995. {PMLR},
  2020{\natexlab{a}}.
\newblock URL \url{http://proceedings.mlr.press/v119/wang20p.html}.

\bibitem[Wang et~al.(2020{\natexlab{b}})Wang, Tsvetkov, and
  Neubig]{wang-etal-2020-balancing}
Wang, X., Tsvetkov, Y., and Neubig, G.
\newblock Balancing training for multilingual neural machine translation.
\newblock In Jurafsky, D., Chai, J., Schluter, N., and Tetreault, J. (eds.),
  \emph{Proceedings of the 58th Annual Meeting of the Association for
  Computational Linguistics}, pp.\  8526--8537, Online, July
  2020{\natexlab{b}}. Association for Computational Linguistics.
\newblock \doi{10.18653/v1/2020.acl-main.754}.
\newblock URL \url{https://aclanthology.org/2020.acl-main.754/}.

\bibitem[Williams(1992)]{DBLP:journals/ml/Williams92}
Williams, R.~J.
\newblock Simple statistical gradient-following algorithms for connectionist
  reinforcement learning.
\newblock \emph{Mach. Learn.}, 8:\penalty0 229--256, 1992.
\newblock \doi{10.1007/BF00992696}.
\newblock URL \url{https://doi.org/10.1007/BF00992696}.

\bibitem[Xia et~al.(2024)Xia, Malladi, Gururangan, Arora, and Chen]{less}
Xia, M., Malladi, S., Gururangan, S., Arora, S., and Chen, D.
\newblock {LESS:} selecting influential data for targeted instruction tuning.
\newblock In \emph{Forty-first International Conference on Machine Learning,
  {ICML} 2024, Vienna, Austria, July 21-27, 2024}. OpenReview.net, 2024.
\newblock URL \url{https://openreview.net/forum?id=PG5fV50maR}.

\bibitem[Xie et~al.(2023{\natexlab{a}})Xie, Pham, Dong, Du, Liu, Lu, Liang, Le,
  Ma, and Yu]{doremi}
Xie, S.~M., Pham, H., Dong, X., Du, N., Liu, H., Lu, Y., Liang, P., Le, Q.~V.,
  Ma, T., and Yu, A.~W.
\newblock Doremi: Optimizing data mixtures speeds up language model
  pretraining.
\newblock In \emph{Thirty-seventh Conference on Neural Information Processing
  Systems}, 2023{\natexlab{a}}.
\newblock URL \url{https://openreview.net/forum?id=lXuByUeHhd}.

\bibitem[Xie et~al.(2023{\natexlab{b}})Xie, Santurkar, Ma, and Liang]{dsir}
Xie, S.~M., Santurkar, S., Ma, T., and Liang, P.
\newblock Data selection for language models via importance resampling.
\newblock In \emph{Thirty-seventh Conference on Neural Information Processing
  Systems}, 2023{\natexlab{b}}.
\newblock URL \url{https://openreview.net/forum?id=uPSQv0leAu}.

\bibitem[Xie et~al.(2023{\natexlab{c}})Xie, Dong, Wang, Lv, Yao, Gan, Wu, Li,
  Li, Liu, and Ma]{10.1145/3539618.3591874}
Xie, X., Dong, Q., Wang, B., Lv, F., Yao, T., Gan, W., Wu, Z., Li, X., Li, H.,
  Liu, Y., and Ma, J.
\newblock T2ranking: A large-scale chinese benchmark for passage ranking.
\newblock In \emph{Proceedings of the 46th International ACM SIGIR Conference
  on Research and Development in Information Retrieval}, SIGIR '23, pp.\
  2681–2690, New York, NY, USA, 2023{\natexlab{c}}. Association for Computing
  Machinery.
\newblock ISBN 9781450394086.
\newblock \doi{10.1145/3539618.3591874}.
\newblock URL \url{https://doi.org/10.1145/3539618.3591874}.

\bibitem[Yang et~al.(2018)Yang, Qi, Zhang, Bengio, Cohen, Salakhutdinov, and
  Manning]{yang-etal-2018-hotpotqa}
Yang, Z., Qi, P., Zhang, S., Bengio, Y., Cohen, W., Salakhutdinov, R., and
  Manning, C.~D.
\newblock {H}otpot{QA}: A dataset for diverse, explainable multi-hop question
  answering.
\newblock In Riloff, E., Chiang, D., Hockenmaier, J., and Tsujii, J. (eds.),
  \emph{Proceedings of the 2018 Conference on Empirical Methods in Natural
  Language Processing}, pp.\  2369--2380, Brussels, Belgium, October-November
  2018. Association for Computational Linguistics.
\newblock \doi{10.18653/v1/D18-1259}.
\newblock URL \url{https://aclanthology.org/D18-1259/}.

\bibitem[Yu et~al.(2024)Yu, Das, and Xiong]{mates}
Yu, Z., Das, S., and Xiong, C.
\newblock {MATES}: Model-aware data selection for efficient pretraining with
  data influence models.
\newblock In \emph{The Thirty-eighth Annual Conference on Neural Information
  Processing Systems}, 2024.
\newblock URL \url{https://openreview.net/forum?id=6gzPSMUAz2}.

\bibitem[Zhang et~al.(2025)Zhang, Zhong, Zhang, Chai, Wang, Zhuang, Bai,
  Jiantao, Cao, Fan, Yuan, Wang, and He]{harnessingdiversitydataselect}
Zhang, C., Zhong, H., Zhang, K., Chai, C., Wang, R., Zhuang, X., Bai, T.,
  Jiantao, Q., Cao, L., Fan, J., Yuan, Y., Wang, G., and He, C.
\newblock Harnessing diversity for important data selection in pretraining
  large language models.
\newblock In \emph{The Thirteenth International Conference on Learning
  Representations}, 2025.
\newblock URL \url{https://openreview.net/forum?id=bMC1t7eLRc}.

\bibitem[Zhang et~al.(2018)Zhang, Zhang, Wang, Guo, and
  Liu]{DBLP:journals/access/ZhangZWGL18}
Zhang, S., Zhang, X., Wang, H., Guo, L., and Liu, S.
\newblock Multi-scale attentive interaction networks for chinese medical
  question answer selection.
\newblock \emph{{IEEE} Access}, 6:\penalty0 74061--74071, 2018.
\newblock \doi{10.1109/ACCESS.2018.2883637}.
\newblock URL \url{https://doi.org/10.1109/ACCESS.2018.2883637}.

\bibitem[Zhang et~al.(2021)Zhang, Ma, Shi, and Lin]{zhang-etal-2021-mr}
Zhang, X., Ma, X., Shi, P., and Lin, J.
\newblock Mr. {T}y{D}i: A multi-lingual benchmark for dense retrieval.
\newblock In Ataman, D., Birch, A., Conneau, A., Firat, O., Ruder, S., and
  Sahin, G.~G. (eds.), \emph{Proceedings of the 1st Workshop on Multilingual
  Representation Learning}, pp.\  127--137, Punta Cana, Dominican Republic,
  November 2021. Association for Computational Linguistics.
\newblock \doi{10.18653/v1/2021.mrl-1.12}.
\newblock URL \url{https://aclanthology.org/2021.mrl-1.12/}.

\bibitem[Zhang et~al.(2023)Zhang, Thakur, Ogundepo, Kamalloo, Alfonso-Hermelo,
  Li, Liu, Rezagholizadeh, and Lin]{zhang-etal-2023-miracl}
Zhang, X., Thakur, N., Ogundepo, O., Kamalloo, E., Alfonso-Hermelo, D., Li, X.,
  Liu, Q., Rezagholizadeh, M., and Lin, J.
\newblock {MIRACL}: A multilingual retrieval dataset covering 18 diverse
  languages.
\newblock \emph{Transactions of the Association for Computational Linguistics},
  11:\penalty0 1114--1131, 2023.
\newblock \doi{10.1162/tacl_a_00595}.
\newblock URL \url{https://aclanthology.org/2023.tacl-1.63/}.

\bibitem[Zhou et~al.(2025)Zhou, Liu, Ma, Zhang, Yuan, Liu, You, and
  Yang]{davir}
Zhou, H., Liu, T., Ma, Q., Zhang, Y., Yuan, J., Liu, P., You, Y., and Yang, H.
\newblock {D}av{IR}: Data selection via implicit reward for large language
  models.
\newblock In Che, W., Nabende, J., Shutova, E., and Pilehvar, M.~T. (eds.),
  \emph{Proceedings of the 63rd Annual Meeting of the Association for
  Computational Linguistics (Volume 1: Long Papers)}, pp.\  9220--9237, Vienna,
  Austria, July 2025. Association for Computational Linguistics.
\newblock ISBN 979-8-89176-251-0.
\newblock \doi{10.18653/v1/2025.acl-long.452}.
\newblock URL \url{https://aclanthology.org/2025.acl-long.452/}.

\bibitem[Zhou et~al.(2024)Zhou, Shi, Song, Yang, Jin, Guo, and
  Li]{DBLP:journals/corr/abs-2406-04614}
Zhou, Z., Shi, J., Song, P., Yang, X., Jin, Y., Guo, L., and Li, Y.
\newblock Lawgpt: {A} chinese legal knowledge-enhanced large language model.
\newblock \emph{CoRR}, abs/2406.04614, 2024.
\newblock \doi{10.48550/ARXIV.2406.04614}.
\newblock URL \url{https://doi.org/10.48550/arXiv.2406.04614}.

\end{thebibliography}


\clearpage

\onecolumn

\appendix

\section{Hyperparameters}
\label{sec:appendix-hyperparameters}

\subsection{BEIR}

For our experiments, we initialize our bi-encoder with \textit{roberta-base} model, we set the initial learning rate to 2e-5, and train with mixed-precision (BF16) enabled. We used an in-batch negative sampling strategy with a temperature of 0.02, normalizing representations before contrastive scoring. Both training and evaluation batch sizes were 256 examples on a single NVIDIA-A100 80GB GPU. For scorer updates, we also use a batch size of 256. We use a linear learning rate decay with a warmup of 250 steps, and trained for a total of 7,000 steps. We use the standard InfoNCE \cite{infonce} as our loss and our influence metric $\mathcal{M}$. Queries and passages were truncated to maximum lengths of 64 and 256 tokens, respectively, with CLS-token pooling for sentence embeddings. We further update our scorer every 50 steps after a 50-step scorer warmup (we don't train the scorer during the warmup period). For reptile updates, we use $\alpha$ based on the current learning rate $\eta_t$ of the learning rate schedule. We do checkpoint selection by selecting the best checkpoint on the dev set and report the corresponding numbers on the final test set. For DoReMi, we train our reference models using $\tau=1$ sampling and proxy models starting with $\tau=\infty$ initialization. For DoGE, too, we train our proxy models starting with $\tau=\infty$ initialization. For CRISP, we limit the clusters to $32^x$: $x \in {1,2}$ since the size of the total dataset is only 800k instances and use \textit{all-MiniLM-L6-v2} embedding model for clustering.

\subsection{Sentence Transformers Embedding}

We initialize our bi-encoder with \textit{nreimers/MiniLM-L6-H384-uncased} model, we set the initial learning rate to 2e-5, and train with mixed-precision (BF16) enabled. We used an in-batch and cross-device negative sampling strategy with a temperature of 0.02, normalizing representations before contrastive scoring. Training batch sizes are set to 2000, and evaluation batch sizes are 256 examples per GPU and use 8 NVIDIA-A100 80GB GPUs. We use the standard InfoNCE \cite{infonce} as our loss and our influence metric $\mathcal{M}$. We also use 1-hard negative when using MSMarco in the training set. For scorer updates, we use a batch size of 256 per dataset. We use a linear learning rate decay with a warmup of 1000 steps, and trained for a total of 150k steps. Queries and passages were truncated to maximum lengths of 64 and 256 tokens, respectively, with CLS-token pooling for sentence embeddings. We further update our scorer every 250 steps after a 500-step scorer warmup (we don't train the scorer during the warmup period). For reptile updates, we use $\alpha$ based on the current learning rate $\eta_t$ of the learning rate schedule. For DoReMi, we train our reference models using $\tau=\infty$ sampling when comparing for uniform initialization $\tau=\text{expert}$ for expert initialization and proxy models starting with $\tau=\infty$ initialization. For DoGE, too, we train our proxy models starting with $\tau=\infty$ initialization sampling when comparing for uniform initialization $\tau=\text{expert}$ for expert initialization. For CRISP, we limit the clusters to $32^x$: $x \in {1,2,3}$ since the size of the total dataset is only 440M instances, and use \textit{all-MiniLM-L6-v2} embedding model for clustering.

\subsection{Multilingual Long Document Retrieval}

We initialize our bi-encoder with \textit{BAAI/bge-m3-unsupervised} model, we set the initial learning rate to 2e-5, and train with mixed-precision (BF16) enabled. We used an in-batch negative, cross-device negatives, and 8 hard negatives during training with a temperature of 0.02, normalizing representations before contrastive scoring. We use the \textit{bge-m3-kd-distil} loss by BGE-M3 as our training loss and InfoNCE with hard negatives as our influence metric $\mathcal{M}$. Training batch sizes are set to 5, and evaluation batch sizes are 5 examples per GPU and use 8 NVIDIA-A100 80GB GPUs. For scorer updates, we use a batch size of 4 per dataset. We use a linear learning rate decay with a warmup of 1000 steps, and trained for a total of 10k steps. Queries and passages were truncated to maximum lengths of 512 and 8192 tokens, respectively, with CLS-token pooling for sentence embeddings. We further update our scorer every 250 steps after a 500-step scorer warmup (we don't train the scorer during the warmup period). We employ subsampling as detailed in Section \ref{subsec:opt_compute_scalability} with $k=8$. We experiment with both enabling and disabling Reptile updates, and observe better performance when Reptile updates are turned off in the MLDR-13 setting. We use reptile update $\alpha$ based on the current learning rate $\eta_t$ of the learning rate schedule. For DoReMi, we train our reference models using $\tau=\infty$ sampling and proxy models starting with $\tau=\infty$ initialization. For DoGE, too, we train our proxy models starting with $\tau=\infty$ initialization. For CRISP, we limit the clusters to $32^x$: $x \in {1,2}$ per language since the size of the total dataset is only 1.5M instances and use \textit{paraphrase-multilingual-MiniLM-L12-v2} embedding model for clustering.

\section{Datasets}
\label{sec:appendix-datasets}

\subsection{BEIR}

\begin{table}[ht]
  \centering
  \begin{minipage}[t]{0.48\textwidth}
    \centering
    \caption{BEIR Train Datasets}
    \label{tab:beir-train}
    \begin{tabular}{@{}lr@{}}
      \toprule
      Dataset    & \#Qrels   \\
      \midrule
      MSMARCO    & 499,184 \\
      NFCorpus   &   2,590 \\
      NQ         & 100,231 \\
      HotpotQA   &  85,000 \\
      FiQA       &   5,500 \\
      Fever      & 109,810 \\
      SciFact    &     809 \\
      \midrule
      \textbf{Total} & \textbf{803,124} \\
      \bottomrule
    \end{tabular}
  \end{minipage}%
  \hfill
  \begin{minipage}[t]{0.48\textwidth}
    \centering
    \caption{BEIR Dev Datasets}
    \label{tab:beir-dev}
    \begin{tabular}{@{}lr@{}}
      \toprule
      Dataset    & \#Qrels   \\
      \midrule
      HotpotQA   & 10,894  \\
      FiQA       &  1,238  \\
      Quora      &  7,626  \\
      DBpedia    &  5,673  \\
      Fever      &  8,079  \\
      \midrule
      \textbf{Total} & \textbf{40,947} \\
      \bottomrule
    \end{tabular}
  \end{minipage}
\end{table}

\subsection{Sentence Transformers Embedding Dataset}

The all‑MiniLM‑L6‑v2 model was fine‐tuned with a self‑supervised contrastive objective over a concatenation of diverse, publicly available sentence‑pair corpora. These include user‑generated content such as paired Reddit comments and Q\&A threads from Stack Exchange and Yahoo Answers; scientific citation pairs drawn from the S2ORC and SPECTER datasets; question answering benchmarks like PAQ, MSMARCO, Natural-Questions, SearchQA, SQuAD 2.0, and TriviaQA; paraphrase and duplicate‑question collections from WikiAnswers, Quora Question Triplets, and the AllNLI (SNLI+MultiNLI) corpus; multimodal captions from COCO and Flickr30k; code‐search examples; and specialized text‐compression and instructional corpora such as Simple Wikipedia, Wikihow, Altlex, and explicit sentence‐compression datasets. Full dataset statistics are provided in \cref{tab:senttrans-init}.

\begin{table*}[ht]
\centering
\resizebox{0.95\textwidth}{!}{%
\begin{tabular}{|l|r|c|c|c|}
\hline
\textbf{Dataset} & \textbf{Count} & \textbf{Proportional Sampling \%} & \textbf{Weight} & \textbf{Expert Sampling \%} \\
\hline
S2ORC Citation pairs (Abstracts) & 116,288,806 & 26.42\% & 123 & 3.07\% \\
WikiAnswers Duplicate question pairs & 77,427,422 & 17.60\% & 123 & 3.07\% \\
Amazon QA Pairs & 2,448,839 & 0.56\% & 247 & 6.16\% \\
PAQ (Question, Answer) pairs & 64,371,441 & 14.63\% & 123 & 3.07\% \\
S2ORC Citation pairs (Titles) & 52,603,982 & 11.96\% & 123 & 3.07\% \\
S2ORC (Title, Abstract) & 41,769,185 & 9.49\% & 123 & 3.07\% \\
Stack Exchange (Title, Body) pairs & 25,316,456 & 5.75\% & 565 & 14.09\% \\
Stack Exchange (Title+Body, Answer) pairs & 21,396,559 & 4.86\% & 17 & 0.42\% \\
Stack Exchange (Title, Answer) pairs & 21,396,559 & 4.86\% & 373 & 9.31\% \\
Stack Exchange Math & 2,218,989 & 0.50\% & 166 & 4.14\% \\
MS MARCO triplets & 9,144,553 & 2.08\% & 247 & 6.16\% \\
GOOAQ: Open QA with Diverse Answer Types & 3,012,496 & 0.68\% & 247 & 6.16\% \\
Yahoo Answers (Title, Answer) & 1,198,260 & 0.27\% & 247 & 6.16\% \\
COCO Image captions & 828,395 & 0.19\% & 1 & 0.02\% \\
SPECTER citation triplets & 684,100 & 0.16\% & 84 & 2.10\% \\
Yahoo Answers (Question, Answer) & 681,164 & 0.15\% & 169 & 4.21\% \\
Yahoo Answers (Title, Question) & 659,896 & 0.15\% & 163 & 4.07\% \\
SearchQA & 582,261 & 0.13\% & 144 & 3.59\% \\
Eli5 & 325,475 & 0.07\% & 81 & 2.02\% \\
Flickr 30k & 317,695 & 0.07\% & 1 & 0.02\% \\
Stack Exchange Duplicate questions (titles) & 304,525 & 0.07\% & 26 & 0.65\% \\
AllNLI (SNLI and MultiNLI) & 277,230 & 0.06\% & 69 & 1.72\% \\
Stack Exchange Duplicate questions (bodies) & 250,519 & 0.06\% & 21 & 0.52\% \\
Stack Exchange Duplicate questions (titles+bodies) & 250,460 & 0.06\% & 21 & 0.52\% \\
Sentence Compression & 180,000 & 0.04\% & 45 & 1.12\% \\
Wikihow & 128,542 & 0.03\% & 32 & 0.80\% \\
Altlex & 112,696 & 0.03\% & 28 & 0.70\% \\
Quora Question Triplets & 103,663 & 0.02\% & 26 & 0.65\% \\
Simple Wikipedia & 102,225 & 0.02\% & 26 & 0.65\% \\
Natural Questions (NQ) & 100,231 & 0.02\% & 25 & 0.62\% \\
SQuAD2.0 & 87,599 & 0.02\% & 22 & 0.55\% \\
TriviaQA & 73,346 & 0.02\% & 19 & 0.47\% \\
\hline
\textbf{Total} & \textbf{440,096,511} & \textbf{100.00\%} & \textbf{4009} & \textbf{100.00\%} \\
\hline
\end{tabular}
}
\caption{Training data provided by sentence transformers \emph{all-MiniLM-L6-v2} showing dataset sizes, hand-picked weights, and normalized sampling percentages.}
\label{tab:senttrans-init}
\end{table*}

\subsection{BGE-m3 Multilingual}

For English, bge-m3 fine-tuning dataset includes 8 datasets, including HotpotQA \cite{yang-etal-2018-hotpotqa}, TriviaQA \cite{joshi-etal-2017-triviaqa}, NQ \cite{kwiatkowski-etal-2019-natural}, MS MARCO \cite{msmarco}, COLIEE \cite{10.1007/978-3-031-29168-5_4}, PubMedQA \cite{jin-etal-2019-pubmedqa}, SQuAD \cite{rajpurkar-etal-2016-squad}, and SimCSE \cite{gao-etal-2021-simcse}. For Chinese, it includes 7 datasets, DuReader \cite{he-etal-2018-dureader}, mMARCO-ZH \cite{DBLP:journals/corr/abs-2108-13897}, T2-Ranking \cite{10.1145/3539618.3591874}, LawGPT \cite{DBLP:journals/corr/abs-2406-04614}, CMedQAv2 \cite{DBLP:journals/access/ZhangZWGL18}, NLIzh2, and LeCaRDv2 \cite{10.1145/3626772.3657887}. It also includes training data for other languages from Mr.
Tydi \cite{zhang-etal-2021-mr}, MIRACL \cite{zhang-etal-2023-miracl} and train sets of MLDR. Full dataset statistics are provided in \cref{tab:bgem3-init}.

\begin{table}[ht]
\centering
\resizebox{0.25\textwidth}{!}{%
\begin{tabular}{|l|r|}
\hline
\textbf{Language} & \textbf{Sampling (\%)} \\
\hline
Swahili    & 0.588  \\
Farsi      & 0.588  \\
Finnish    & 0.294  \\
Indonesian & 0.294  \\
French     & 1.176  \\
German     & 1.176  \\
Korean     & 1.176  \\
Spanish    & 1.176  \\
Italian    & 1.176  \\
Portuguese & 1.176  \\
Japanese   & 1.176  \\
Bengali    & 0.294  \\
Telugu     & 0.294  \\
Thai       & 1.176  \\
Russian    & 2.353  \\
Hindi      & 1.176  \\
Arabic     & 2.353  \\
Chinese    & 23.529 \\
English    & 58.824 \\
\hline
\end{tabular}
}
\caption{Language wise initialization probabilities for \emph{bge-m3-dense} training.}
\label{tab:bge-m3-langwiseinit}
\end{table}

\begin{table*}[ht]
\centering
\begin{minipage}[t]{0.4\textwidth}
    \centering
    \resizebox{\textwidth}{!}{%
    \begin{tabular}{|l|r|c|}
    \hline
    \textbf{Dataset} & \textbf{Lines} & \textbf{Sampling \%} \\
    \hline
    MSMarco              & 485,905 & 30.95\% \\
    MIRACL/fr                    &   1,143 &  0.07\% \\
    MIRACL/zh                    &   1,312 &  0.08\% \\
    MIRACL/es                    &   2,162 &  0.14\% \\
    MIRACL/ja                    &   3,477 &  0.22\% \\
    MIRACL/te                    &   3,452 &  0.22\% \\
    MIRACL/en                    &   2,863 &  0.18\% \\
    MIRACL/id                    &   4,071 &  0.26\% \\
    MIRACL/fa                    &   2,107 &  0.13\% \\
    MIRACL/ko                    &     868 &  0.06\% \\
    MIRACL/fi                    &   2,897 &  0.18\% \\
    MIRACL/th                    &   2,972 &  0.19\% \\
    MIRACL/bn                    &   1,631 &  0.10\% \\
    MIRACL/ru                    &   4,683 &  0.30\% \\
    MIRACL/hi                    &   1,169 &  0.07\% \\
    MIRACL/ar                    &   3,495 &  0.22\% \\
    MIRACL/sw                    &   1,901 &  0.12\% \\
    HotpotQA                     &  84,516 &  5.38\% \\
    mMARCO-zh/chinese            & 100,000 &  6.37\% \\
    NQ                           &  58,568 &  3.73\% \\
    zh\_NLI/LCQMC                &  10,000 &  0.64\% \\
    zh\_NLI/BQ                   &  12,599 &  0.80\% \\
    zh\_NLI/STS-B                &     249 &  0.02\% \\
    zh\_NLI/afqmc                &  10,534 &  0.67\% \\
    zh\_NLI/ATEC                 &  11,325 &  0.72\% \\
    zh\_NLI/QBQTC\_v2            &  10,000 &  0.64\% \\
    zh\_NLI/PAWSX                &  10,000 &  0.64\% \\
    DuReader                     &  80,416 &  5.12\% \\
    cMedQAv2                     &  50,000 &  3.18\% \\
    TriviaQA                     &  60,315 &  3.84\% \\
    \hline
    \end{tabular}%
    }
\end{minipage}
\begin{minipage}[t]{0.4\textwidth}
    \centering
    \resizebox{\textwidth}{!}{%
    \begin{tabular}{|l|r|c|}
    \hline
    \textbf{Dataset} & \textbf{Lines} & \textbf{Sampling \%} \\
    \hline
    Mr.TyDi/finnish              &   6,561 &  0.42\% \\
    Mr.TyDi/bengali              &   1,713 &  0.11\% \\
    Mr.TyDi/russian              &   5,366 &  0.34\% \\
    Mr.TyDi/swahili              &   2,072 &  0.13\% \\
    Mr.TyDi/indonesian           &   4,902 &  0.31\% \\
    Mr.TyDi/arabic               &  12,377 &  0.79\% \\
    Mr.TyDi/english              &   3,547 &  0.23\% \\
    Mr.TyDi/korean               &   1,295 &  0.08\% \\
    Mr.TyDi/japanese             &   3,697 &  0.24\% \\
    Mr.TyDi/telugu               &   3,880 &  0.25\% \\
    Mr.TyDi/thai                 &   3,319 &  0.21\% \\
    T2Ranking                    &  90,467 &  5.76\% \\
    en\_NLI/nli\_for\_simcse      & 274,951 & 17.51\% \\
    Law-Medical/colliee          &     463 &  0.03\% \\
    Law-Medical/law\_gpt         &     500 &  0.03\% \\
    Law-Medical/lecardv2         &     591 &  0.04\% \\
    Law-Medical/pubmed\_qa       &     500 &  0.03\% \\
    MLDR/hi                      &   1,618 &  0.10\% \\
    MLDR/es                      &   2,254 &  0.14\% \\
    MLDR/ru                      &   1,864 &  0.12\% \\
    MLDR/de                      &   1,847 &  0.12\% \\
    MLDR/ja                      &   2,262 &  0.14\% \\
    MLDR/fr                      &   1,608 &  0.10\% \\
    MLDR/ar                      &   1,817 &  0.12\% \\
    MLDR/ko                      &   2,198 &  0.14\% \\
    MLDR/en                      &  10,000 &  0.64\% \\
    MLDR/zh                      &  10,000 &  0.64\% \\
    MLDR/pt                      &   1,845 &  0.12\% \\
    MLDR/it                      &   2,151 &  0.14\% \\
    MLDR/th                      &   1,970 &  0.13\% \\
    SQuAD                        &  87,599 &  5.58\% \\
    \hline
    \textbf{Total}               & \textbf{1,569,864} & \textbf{100.00\%} \\
    \hline
    \end{tabular}%
    }
\end{minipage}
\caption{Training data provided by \emph{bge-m3} showing dataset sizes and normalized sampling percentages.}
\label{tab:bgem3-init}
\end{table*}

\section{Additional results}
\label{sec:appendix-results}

\subsection{BEIR}
\label{app:additional-BEIR}

\begin{figure*}[ht]
  \centering
  \includegraphics[width=0.25\linewidth]{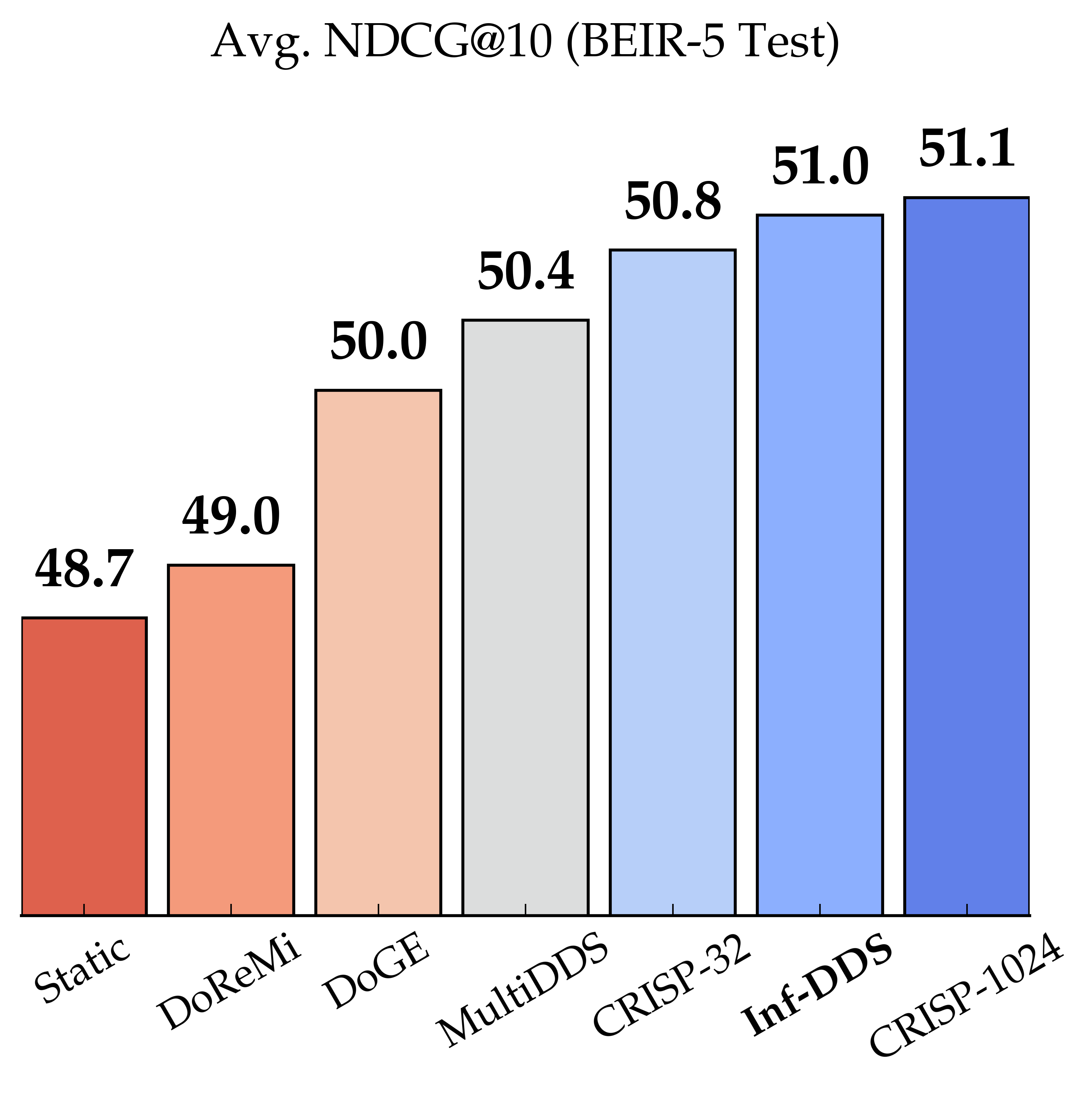}
  \caption{Average NDCG@10 on the BEIR-5 test collection using BEIR-7 training data with $\tau=1$. The scorer is optimized individually on the development sets.}
  \label{fig:beir7beir5individual}
\end{figure*}

To further validate our approach, we conduct individual optimizations on each of the BEIR-5 development sets and report the results in Table~\ref{tab:beir7beir5individual} Figure~\ref{fig:beir7beir5individual}. As shown, \ouralgo{} consistently matches or outperforms static sampling across all datasets. While it does not achieve the highest performance in every domain compared to MultiDDS, \ouralgo{} surpasses proportional sampling by an average margin of \(2.24\) points and even outperforms MultiDDS on average. The corresponding sampling trajectories are provided in the Appendix \ref{sec:appendix-samplinggraphs}. 

\begin{table}[h]
  \centering
  \resizebox{0.5\textwidth}{!}{%
\begin{tabular}{c|cccc}
\toprule
Init.         & \begin{tabular}[c]{@{}c@{}}Static\\ Sampling\end{tabular} & MultiDDS & Inf-DDS & Others \\ 
\hline
$\tau=0.3$$^{\dagger}$    & 57.3 & 54.5 & \textbf{64.9}$^{*}$ & - \\
$\tau=1$      & 48.8 & 53.9 & 54.7 & - \\
$\tau=5$      & 46.5 & 53.2 & 55.2 & - \\
$\tau=\infty$ & 50.1 & 55.8 & 57.5 & - \\ 
\hline
Cooldown      & -     & -     & -     & 50.5 \\
DoReMi        & -     & -     & -     & 48.6 \\
DoGE          & -     & -     & -     & 54.7 \\
CRISP-32      & -     & -     & -     & 59.0 \\
CRISP-1024$^{\ddagger}$    & -     & -     & -     & 62.9 \\
\bottomrule
\end{tabular}
}
\caption{NDCG@10 comparison of sampling strategies with varying initialization temperatures during optimization on the FEVER dev set. $^{\dagger}$ and $^{\ddagger}$ mark the two models compared in the paired significance test, and $^{*}$ indicates a statistically significant difference ($p < 0.05$). Additional results from other baselines are shown in the last column.}
\label{tab:beir7trainfever}
\end{table}

\begin{table*}[ht]
  \centering
  \resizebox{0.9\textwidth}{!}{%
\begin{tabular}{c|c|c|c|c|ccccc}
\toprule
Sampling &
  Init. &
  \begin{tabular}[c]{@{}c@{}}Train \\ Dataset\end{tabular} &
  \begin{tabular}[c]{@{}c@{}}Dev \\ Dataset\end{tabular} &
  \begin{tabular}[c]{@{}c@{}}Average\\ Test\end{tabular} &
  \multicolumn{1}{l}{DBpedia} &
  FEVER &
  \multicolumn{1}{l}{FiQA} &
  HotpotQA &
  Quora \\ \hline
Static &
  $\tau=1$ &
  BEIR-7 &
  - &
  48.7 &
  \textbf{29.5} &
  48.8 &
  29.0 &
  52.4 &
  \textbf{84.0} \\ \cline{1-1}
MultiDDS &
  $\tau=1$ &
  BEIR-7 &
  BEIR-1 &
  50.4 &
  27.4 &
  53.9 &
  \textbf{33.3} &
  \textbf{53.8} &
  83.7 \\ \cline{1-1}
Inf-DDS &
  $\tau=1$ &
  BEIR-7 &
  BEIR-1 &
  \textbf{51.0} &
  \textbf{29.4} &
  54.7 &
  \textbf{33.7} &
  52.9 &
  \textbf{84.1} \\ \hline
DoReMi &
  Proxy &
  BEIR-7 &
  - &
  49.0 &
  27.5 &
  48.6 &
  32.0 &
  53.7 &
  83.1 \\ \cline{1-1}
DoGE &
  Proxy &
  BEIR-7 &
  BEIR-1 &
  50.0 &
  25.1 &
  54.7 &
  32.4 &
  \textbf{54.8} &
  82.8 \\ \cline{1-1}
CRISP-32 &
  Cluster IS &
  BEIR-7 &
  BEIR-1 &
  50.8 &
  27.0 &
  \textbf{59.0} &
  31.3 &
  52.9 &
  83.6 \\ \cline{1-1}
CRISP-1024 &
  Cluster IS &
  BEIR-7 &
  BEIR-1 &
  \textbf{51.1} &
  24.8 &
  \textbf{62.9} &
  31.8 &
  52.6 &
  83.3 \\ \hline
Static &
  $\tau=\infty$ &
  BEIR-7 &
  - &
  48.9 &
  27.2 &
  50.1 &
  31.7 &
  52.8 &
  82.8 \\ \cline{1-1}
MultiDDS &
  $\tau=\infty$ &
  BEIR-7 &
  BEIR-1 &
  49.8 &
  25.6 &
  55.8 &
  31.1 &
  53.1 &
  83.1 \\ \cline{1-1}
\ouralgo{} &
  $\tau=\infty$ &
  BEIR-7 &
  BEIR-1 &
  50.3 &
  26.7 &
  57.5 &
  32.0 &
  52.6 &
  82.7 \\
\bottomrule
\end{tabular}
}
\caption{Average NDCG@10 on BEIR-5 test collection, optimization of scorer is done individually on the development sets.}
\label{tab:beir7beir5individual}
\end{table*}

\begin{table*}[ht]
  \centering
  \resizebox{0.85\textwidth}{!}{%
\begin{tabular}{c|c|c|c|c|ccccc}
\toprule
Sampling &
  Init &
  \begin{tabular}[c]{@{}c@{}}Train \\ Dataset\end{tabular} &
  \begin{tabular}[c]{@{}c@{}}Dev \\ Dataset\end{tabular} &
  \begin{tabular}[c]{@{}c@{}}Avg.\\ Test\end{tabular} &
  Dbpedia &
  Fever &
  Fiqa &
  Hotpotqa &
  Quora \\ \hline
Static   &  &  & - & 49.2 & 28.6 & 58.2 & 26.5 & 48.7 & 84.0 \\ \cline{4-4}
MultiDDS &  &  &   & 48.2 & 25.2 & 58.6 & 23.3 & 54.0 & 80.1 \\
\ouralgo{}$^{\dagger}$ &
  \multirow{-3}{*}{$\tau=0.3$} &
   &
  \multirow{-2}{*}{BEIR-5} &
  50.3 & 27.7$^{*}$ & 69.2$^{*}$ & 23.6 & 48.6 & 82.3$^{*}$ \\ \cline{1-2} \cline{4-10} 
Static   &  &  & - & 48.7 & 29.5 & 48.8 & 29.0 & 52.4 & 84.0 \\ \cline{4-4}
MultiDDS &  &  &   & 49.3 & 25.9 & 59.0 & 29.0 & 50.5 & 82.0 \\
\ouralgo{} &
  \multirow{-3}{*}{$\tau=1$} &
   &
  \multirow{-2}{*}{BEIR-5} &
  48.9 & 29.5 & 50.0 & 29.4 & 51.6 & 84.1 \\ \cline{1-2} \cline{4-10} 
Static   &  &  & - & 48.9 & 27.6 & 48.4 & 32.4 & 53.2 & 83.0 \\ \cline{4-4}
MultiDDS &  &  &   & 47.4 & 24.4 & 51.0 & 30.1 & 49.7 & 81.5 \\
\ouralgo{} &
  \multirow{-3}{*}{$\tau=5$} &
   &
  \multirow{-2}{*}{BEIR-5} &
  49.7 & 28.8 & 55.0 & 31.2 & 51.4 & 82.2 \\ \cline{1-2} \cline{4-10} 
Static   &  &  & - & 48.9 & 27.2 & 50.1 & 31.7 & 52.8 & 82.8 \\ \cline{4-4}
MultiDDS &  &  &   & 47.3 & 22.7 & 52.4 & 30.4 & 50.1 & 81.1 \\
\ouralgo{} &
  \multirow{-3}{*}{$\tau=\infty$} &
  \multirow{-12}{*}{BEIR-7} &
  \multirow{-2}{*}{BEIR-5} &
  50.2 & 26.8 & 59.6 & 32.1 & 50.4 & 82.2 \\ \hline
DoReMi &
  Proxy &
  BEIR-7 &
  - &
  49.0 & 27.5 & 48.6 & 32.0 & 53.7 & 83.1 \\ \cline{1-1}
DoGE$^{\ddagger}$ &
  Proxy &
  BEIR-7 &
  BEIR-5 &
  50.4 & 26.7 & 61.3 & 30.4$^{*}$ & 51.4$^{*}$ & 82.2 \\ \cline{1-1}
CRISP-32 &
  Cluster IS &
  BEIR-7 &
  BEIR-5 &
  50.0 & 26.2 & 57.2 & 31.0 & 52.3 & 83.5 \\ \cline{1-1}
CRISP-1024 &
  Cluster IS &
  BEIR-7 &
  BEIR-5 &
  49.4 & 23.6 & 56.1 & 31.2 & 52.5 & 83.3 \\
\bottomrule
\end{tabular}
}
\caption{Average NDCG@10 on the BEIR-5 test collections. Scorer optimization is performed jointly on the development sets. $^{\dagger}$ and $^{\ddagger}$ denote the two models compared in the paired significance test, and $^{*}$ indicates a statistically significant difference ($p < 0.05$). Additional baselines are listed below.}
\label{tab:beir7beir5joint}
\end{table*}

\subsection{Sentence Transformers Embedding Dataset}
\label{app:additional-st}

\begin{table*}[ht]
  \centering
  \resizebox{0.95\textwidth}{!}{%
\begin{tabular}{c|c|c|c|c|ccccc}
\toprule
Sampling &
  Init. &
  Train Dataset &
  Dev Dataset &
  \begin{tabular}[c]{@{}c@{}}Average\\ Test\end{tabular} &
  DBpedia &
  FEVER &
  FiQA &
  HotpotQA &
  Quora \\ \hline
Static &
  $\tau=\infty$ &
  Sent-Trans &
  - &
  49.0 & 31.7 & 49.0 & 31.8 & 44.6 & 49.0 \\ \cline{1-1}
MultiDDS &
  $\tau=\infty$ &
  Sent-Trans &
  BEIR-5 &
  40.2 & 22.3 & 43.9 & 24.8 & 26.5 & 83.9 \\ \cline{1-1}
InfDDS &
  $\tau=\infty$ &
  Sent-Trans &
  BEIR-5 &
  50.8 & 32.0 & 50.6 & 34.4 & 46.6 & 88.0 \\ \hline
Cooldown &
  $5 \rightarrow 1$ &
  Sent-Trans &
  - &
  47.6 & 28.6 & 49.9 & 33.7 & 38.1 & 87.6 \\ \cline{1-1}
DoReMi &
  Proxy &
  Sent-Trans &
  - &
  51.4 & 31.7 & 52.3 & 36.2 & 49.1 & 87.6 \\ \cline{1-1}
DoGE &
  Proxy &
  Sent-Trans &
  BEIR-5 &
  49.5 & 29.9 & 57.4 & 31.2 & 42.6 & 86.3 \\ \cline{1-1}
CRISP (n=32) &
  Cluster IS &
  Sent-Trans &
  BEIR-5 &
  47.4 & 28.3 & 58.1 & 27.9 & 38.0 & 84.9 \\ \cline{1-1}
CRISP (n=32$^2$) &
  Cluster IS &
  Sent-Trans &
  BEIR-5 &
  45.6 & 26.8 & 54.8 & 26.4 & 35.5 & 84.7 \\ \cline{1-1}
CRISP (n=32$^3$) &
  Cluster IS &
  Sent-Trans &
  BEIR-5 &
  48.5 & 28.4 & 61.9 & 28.5 & 39.5 & 84.2 \\ \hline
Static (all-MiniLM-L6-v2)$^{\ddagger}$ &
  Expert &
  Sent-Trans &
  - &
  51.0 & 32.3 & 51.9 & 36.9$^{*}$ & 46.5$^{*}$ & 87.6 \\ \hline
MultiDDS &
  Expert &
  Sent-Trans &
  BEIR-5 &
  42.5 & 22.8 & 37.2 & 29.2 & 28.3 & 86.3 \\ \cline{1-1}
InfDDS$^{\dagger}$ &
  Expert &
  Sent-Trans &
  BEIR-5 &
  52.0 & 32.4$^{*}$ & 58.3$^{*}$ & 36.1 & 45.7 & 87.4 \\ \cline{1-1}
DoReMi &
  Expert &
  Sent-Trans &
  - &
  52.3 & 32.5 & 56.3 & 36.2 & 48.7 & 87.7 \\ \cline{1-1}
DoGE &
  Expert &
  Sent-Trans &
  BEIR-5 &
  50.5 & 30.0 & 59.6 & 33.3 & 43.1 & 86.4 \\
\bottomrule
\end{tabular}
}
\caption{Average NDCG@10 on BEIR-5 test collection when training \emph{all-MiniLM-L6-v2} using Sentence-Transformers data, optimization of scorer is done jointly on BEIR-5 development sets. $^{\dagger}$ and $^{\ddagger}$ mark the two models compared in the paired significance test, and $^{*}$ indicates a statistically significant difference ($p < 0.05$).}
\label{tab:senttransbeir5}
\end{table*}

Here in Table \ref{tab:beir7beir5individual} we report Average NDCG@10 numbers on BEIR‑5 test collection when training \emph{all-MiniLM-L6-v2} using Sentence-Transformers data, when optimization of scorer is done jointly on the BEIR-5 development sets.

\subsection{MLDR}
\label{app:additional-MLDR}

Here in Table \ref{tab:mldr13} we report Average NDCG@10 numbers on MLDR-13 test set when training with BGE-M3 data, when optimization of scorer is done jointly on the MLDR-13 development sets.

\begin{table*}[]
  \centering
  \resizebox{\textwidth}{!}{%
\begin{tabular}{c|c|c|ccccccccccccc}
\toprule
\begin{tabular}[c]{@{}c@{}}Model/\\ Sampling\end{tabular} &
  \begin{tabular}[c]{@{}c@{}}Dev \\ Dataset\end{tabular} &
  \begin{tabular}[c]{@{}c@{}}Avg.\\ Test\end{tabular} &
  ar & de & en & es & fr & hi & it & ja & ko & pt & ru & th & zh \\ \hline
\begin{tabular}[c]{@{}c@{}}bge-m3\\ unsup.\end{tabular} &
  - & 37.0 & 30.8 & 38.4 & 34.2 & 61.4 & 53.1 & 23.9 & 43.7 & 33.3 & 24.9 & 59.7 & 45.8 & 18.6 & 13.7 \\ \hline
bge-m3-dense &
  - & 52.5 & 47.6 & 46.1 & 48.9 & 74.8 & 73.8 & 40.7 & 62.7 & 50.9 & 42.9 & 74.4 & 59.5 & 33.6 & 26.0 \\
bge-m3-dense$^{\diamond}$ &
  - & 52.5 & 48.4 & 46.7 & 46.7 & 76.3 & 74.3 & 40.0 & 61.7 & 49.1 & 41.0 & 74.3 & 60.8 & 36.0 & 26.7 \\
MultiDDS &
  MLDR-13 & 56.7 & 56.0 & 53.9 & 50.0 & 79.1 & 76.3 & 44.1 & 66.0 & 56.6 & 49.3 & 78.6 & 62.3 & 37.3 & 27.3 \\
Inf-DDS$^{\dagger}$ &
  MLDR-13 & 57.5 & 58.3$^{*}$ & 54.4 & 48.5 & 80.6$^{*}$ & 77.9$^{*}$ & 43.9 & 66.1 & 56.7$^{*}$ & 49.9$^{*}$ & 79.5 & 65.7$^{*}$ & 39.0 & 26.6 \\ \hline
Cooldown &
  - & 52.7 & 51.5 & 50.2 & 48.5 & 75.6 & 72.6 & 33.1 & 61.7 & 48.4 & 41.3 & 75.5 & 61.4 & 37.8 & 27.7 \\
DoReMi$^{\ddagger}$ &
  - & 57.2 & 54.8 & 54.4 & 50.7$^{*}$ & 79.1 & 77.1 & 44.5$^{*}$ & 66.6$^{*}$ & 54.1 & 48.0 & 82.3$^{*}$ & 64.5 & 39.9$^{*}$ & 27.8$^{*}$ \\
DoGE (Iter 2) &
  MLDR-13 & 49.6 & 44.9 & 47.0 & 47.1 & 72.7 & 70.5 & 34.6 & 58.5 & 44.4 & 37.0 & 73.3 & 56.7 & 33.2 & 25.0 \\
CRISP (n=32/lang) &
  MLDR-13 & 57.1 & 56.7 & 53.3 & 52.6 & 79.5 & 75.6 & 44.0 & 66.6 & 55.4 & 48.8 & 79.3 & 64.7 & 36.7 & 28.9 \\
CRISP (n=32\textsuperscript{2}/lang) &
  MLDR-13 & 56.6 & 53.1 & 53.9 & 52.0 & 79.8 & 75.5 & 44.9 & 65.5 & 54.5 & 48.6 & 78.0 & 62.8 & 38.5 & 28.5 \\ 
\bottomrule
\end{tabular}
}
\caption{Average NDCG@10 scores on the MLDR-13 test set across 13 languages. $^{\diamond}$ denotes reproduced results, $^{\dagger}$ and $^{\ddagger}$ mark the two models compared in the paired significance test, and $^{*}$ indicates a statistically significant difference ($p < 0.05$).}
\label{tab:mldr13}
\end{table*}

\subsection{Ablation}
\label{app:additional-ablation}

\paragraph{Effect of reptile updates:} We perform an ablation study to determine whether the observed performance gains arise solely from dynamically updating the sampling distribution or whether the meta‐learning component of our updates also contributes. Table \ref{tab:ablationreptile} compares downstream performance with Reptile updates enabled versus disabled. We observe only minor differences in performance, which suggests that most of the improvement can be attributed to dynamic sampling. However, because the Reptile meta‐update reuses existing computations, it remains valuable for reducing overall computational overhead.

\begin{table*}[ht]
  \centering
  \resizebox{0.5\textwidth}{!}{%
\begin{tabular}{c|c|c|ccc}
\hline
\begin{tabular}[c]{@{}c@{}}Reptile \\ Update\end{tabular} &
  \begin{tabular}[c]{@{}c@{}}Train \\ Dataset\end{tabular} &
  \begin{tabular}[c]{@{}c@{}}Dev \\ Dataset\end{tabular} &
  \begin{tabular}[c]{@{}c@{}}Fever\\ Test\end{tabular} &
  \begin{tabular}[c]{@{}c@{}}Fiqa\\ Test\end{tabular} &
  \begin{tabular}[c]{@{}c@{}}HotpotQA\\ Test\end{tabular} \\ \hline
On &
  \multirow{2}{*}{BEIR-7} &
  \multirow{2}{*}{BEIR-1} &
  54.7 &
  33.7 &
  52.9 \\
Off &
   & 
   & 56.5 
   & 28.8
   & 52.7
   \\ \hline
\end{tabular}
}
\caption{Average NDCG@10 on FEVER, FiQA, and HotpotQA test collection, optimization of scorer is done individually on the development sets without reptile updates.}
\label{tab:ablationreptile}
\end{table*}

\begin{table*}[t]
\centering
\small
\setlength{\tabcolsep}{10pt}
\renewcommand{\arraystretch}{1.2}
  \resizebox{0.3\textwidth}{!}{%
\begin{tabular}{c c c}
\toprule
\textbf{Update steps $l$} & \textbf{FEVER} & \textbf{FiQA} \\
\midrule
0  & 57.3 & 29.0 \\
1  & 59.4 & 30.9 \\
3  & 64.9 & 33.7 \\
5  & 62.4 & 33.4 \\
10 & 62.9 & 31.2 \\
20 & 60.6 & 31.4 \\
\bottomrule
\end{tabular}
}
\caption{Effect of the number of update steps $l$ on NDCG@10 when optimizing separately for FEVER and FiQA.}
\label{tab:update_steps}
\end{table*}

\begin{table}
\centering
\small
\setlength{\tabcolsep}{5pt}
\renewcommand{\arraystretch}{1.15}
  \resizebox{0.5\textwidth}{!}{%
\begin{tabular}{lccccccc}
\toprule
Language & mix1 & mix2 & mix3 & mix4 & mix5 & Train & Mix Avg. \\
\midrule
ar & 35.7 & 35.7 & 35.9 & 35.2 & 34.7 & 36.2 & 35.44 $\pm$ 0.5 \\
de & 34.2 & 34.0 & 33.7 & 34.4 & 34.1 & 34.5 & 34.08 $\pm$ 0.3 \\
en & 29.5 & 30.2 & 29.4 & 30.7 & 29.3 & 29.6 & 29.82 $\pm$ 0.6 \\
es & 58.7 & 59.8 & 60.7 & 59.4 & 59.5 & 60.3 & 59.62 $\pm$ 0.7 \\
fr & 56.7 & 57.0 & 58.4 & 58.2 & 56.6 & 57.2 & 57.38 $\pm$ 0.9 \\
hi & 26.7 & 25.6 & 25.0 & 23.7 & 23.8 & 25.1 & 24.96 $\pm$ 1.3 \\
it & 49.5 & 49.6 & 48.6 & 49.5 & 48.2 & 49.1 & 49.08 $\pm$ 0.6 \\
ja & 39.4 & 38.8 & 38.4 & 39.8 & 39.0 & 38.5 & 39.08 $\pm$ 0.5 \\
ko & 34.7 & 32.6 & 32.7 & 32.9 & 31.7 & 32.6 & 32.92 $\pm$ 1.1 \\
pt & 61.3 & 59.4 & 61.4 & 60.8 & 60.0 & 61.7 & 60.58 $\pm$ 0.9 \\
ru & 49.8 & 49.6 & 50.6 & 49.8 & 49.3 & 51.4 & 49.82 $\pm$ 0.5 \\
th & 26.4 & 25.0 & 24.6 & 25.2 & 24.3 & 25.7 & 25.10 $\pm$ 0.8 \\
zh & 17.6 & 16.8 & 17.5 & 17.4 & 17.8 & 17.5 & 17.42 $\pm$ 0.4 \\
\midrule
Avg. & 40.0 & 39.5 & 39.8 & 39.8 & 39.1 & 40.0 & 39.6 $\pm$ 0.3 \\
\bottomrule
\end{tabular}
}
\caption{Testing Dev/Test leakage on \emph{bge-m3-retromae} via repeated sampling over the mixture of train and dev splits of MLDR, with evaluation on the MLDR-13 test set. We report NDCG@10 across five mixed-data runs, along with the mix mean and standard deviation.}
\label{tab:mldr_test_leakage}
\end{table}

\begin{table}
\centering
\small
\setlength{\tabcolsep}{5pt}
\renewcommand{\arraystretch}{1.15}
\resizebox{0.5\textwidth}{!}{%
\begin{tabular}{lccccccc}
\toprule
Dataset & mix1 & mix2 & mix3 & mix4 & mix5 & Train & Mix Avg. \\
\midrule
FEVER     & 78.1 & 77.7 & 78.5 & 77.5 & 78.3 & 76.6 & 78.02 $\pm$ 0.4 \\
FiQA      & 30.7 & 30.6 & 30.8 & 30.4 & 30.8 & 31.0 & 30.66 $\pm$ 0.2 \\
HotpotQA  & 30.4 & 30.1 & 30.3 & 30.4 & 30.0 & 30.0 & 30.24 $\pm$ 0.2 \\
\midrule
Avg.      & 46.4 & 46.1 & 46.5 & 46.1 & 46.4 & 45.9 & 46.3 $\pm$ 0.2 \\
\bottomrule
\end{tabular}
}
\caption{Testing Dev/Test leakage on \emph{roberta-base} via repeated sampling over the mixture of BEIR train and development splits, with evaluation on the BEIR test collections. We report NDCG@10 across five mixed-data runs, along with the mix mean and standard deviation.}
\label{tab:beir_test_leakage}
\end{table}

\section{Sampling trajectories}
\label{sec:appendix-samplinggraphs}

\begin{figure*}[ht]
  \centering
  \includegraphics[width=0.9\linewidth]{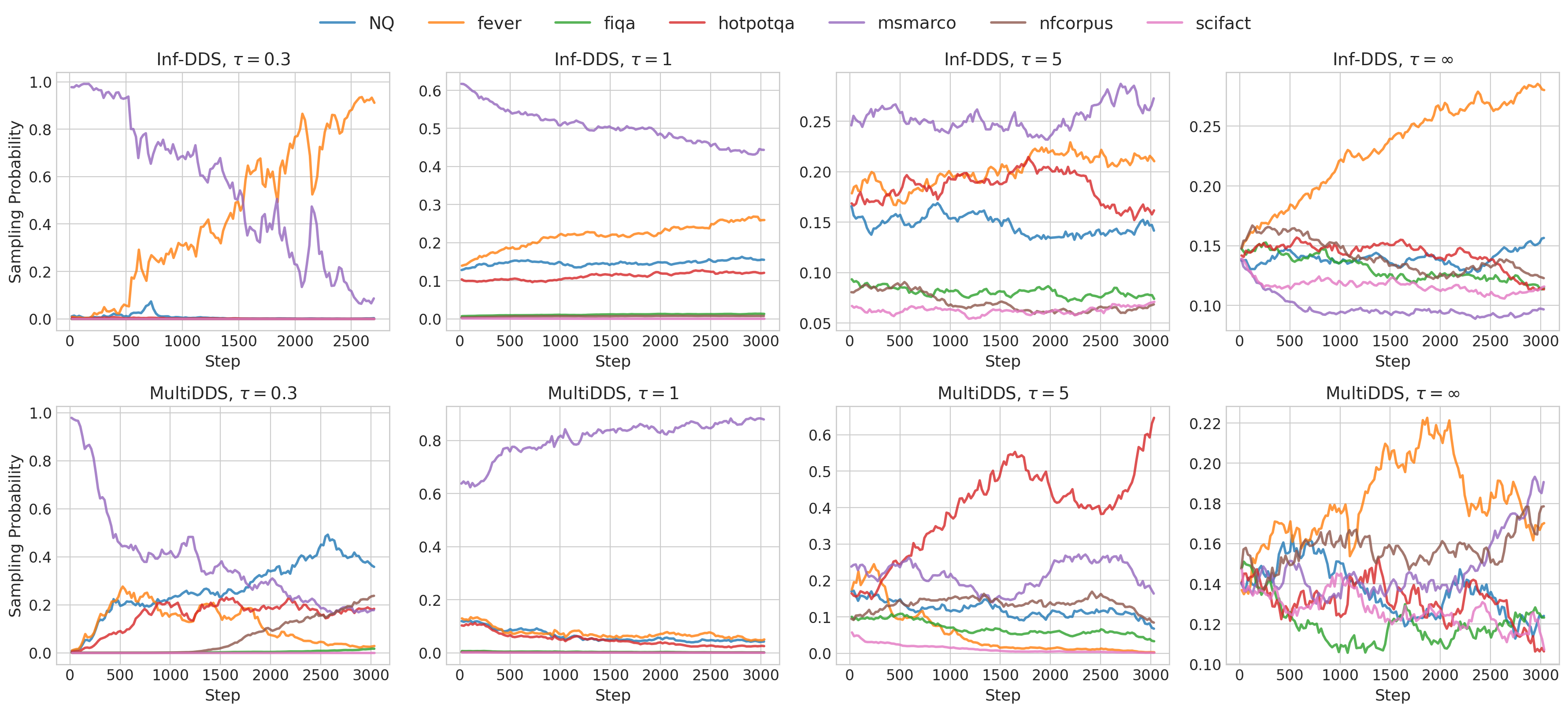}
\caption{Sampling probability trajectories of MultiDDS and \ouralgo{} with varying initialization temperatures during optimization on the FEVER development set. The \textcolor[HTML]{DF672A}{orange} curve denotes the FEVER training set sampling trajectory.}
  \label{fig:plotfeverall}
\end{figure*}

\begin{figure*}[ht]
  \centering
  \includegraphics[width=0.9\linewidth]{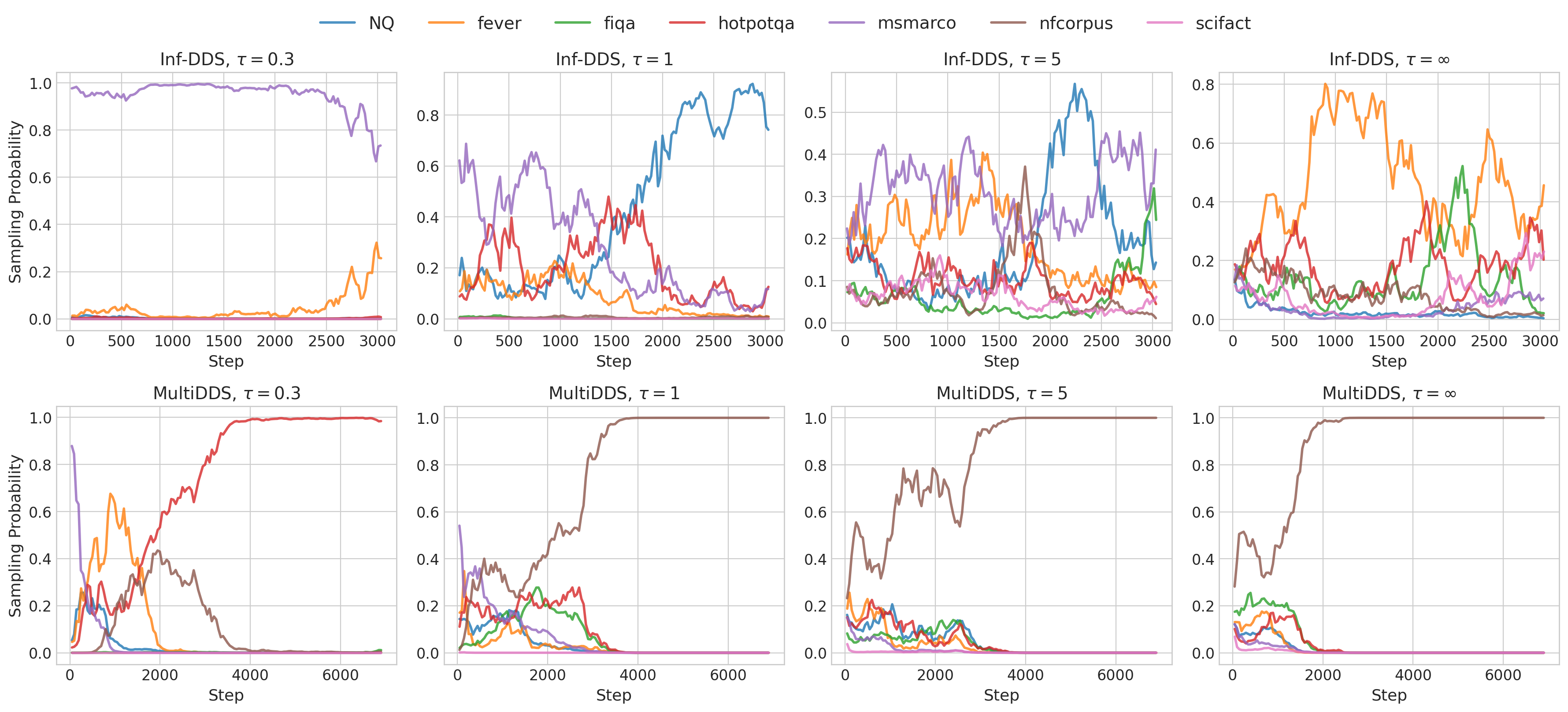}
\caption{Sampling probability trajectories of MultiDDS and \ouralgo{} with varying initialization temperatures during joint optimization on the BEIR-5 development set (DBpedia, FEVER, FiQA, HotpotQA, Quora). The \textcolor{Brown}{brown} curve represents the sampling trajectory for the NFCorpus training set, which is aggressively upsampled by MultiDDS, resulting in degraded overall performance, as shown in Table~\ref{tab:beir7beir5joint}.}
  \label{fig:plotall5}
\end{figure*}

\begin{figure*}[ht]
  \centering
  \includegraphics[width=0.9\linewidth]{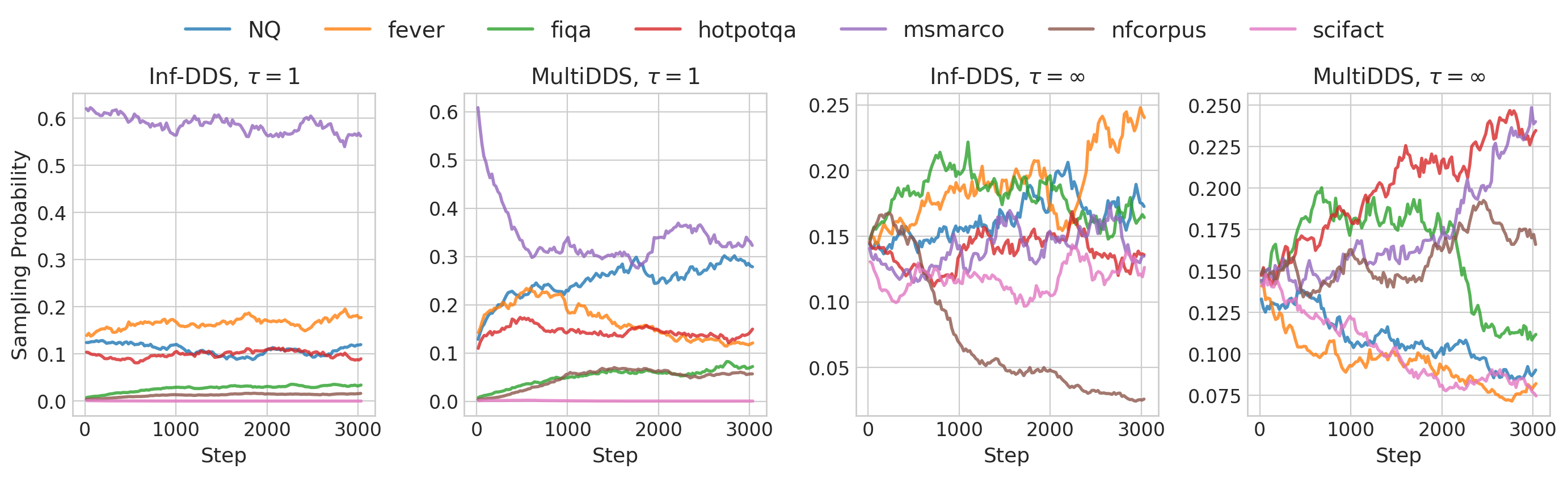}
\caption{Sampling probability trajectories of MultiDDS and \ouralgo{} with varying initialization temperatures during optimization on the FiQA development set. The \textcolor{ForestGreen}{green} curve denotes the FiQA training set sampling trajectory. Compared to MultiDDS, which upsamples FiQA due to gradient similarity, \ouralgo{} upsamples datasets like MSMarco and FEVER that are more relevant for performance gains as seen in Table \ref{tab:beir7beir5individual}.}
  \label{fig:plotfiqa}
\end{figure*}

\begin{figure*}[ht]
  \centering
  \includegraphics[width=0.9\linewidth]{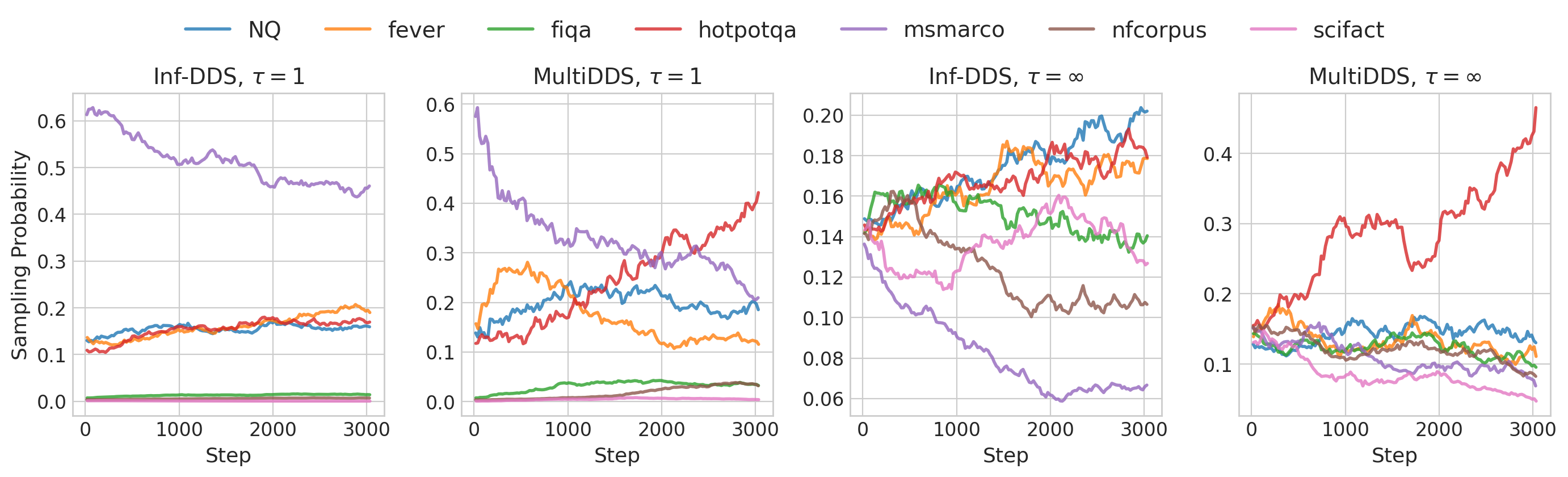}
\caption{Sampling probability trajectories of MultiDDS and \ouralgo{} with varying initialization temperatures during joint optimization on the HotpotQA development set. The \textcolor{Red}{red} curve represents the sampling trajectory for the HotpotQA training set, which is being upsampled more by MultiDDS, leading to a better performance as seen in Table \ref{tab:beir7beir5individual}.}
  \label{fig:plothotpotqa}
\end{figure*}

\begin{figure*}[ht]
  \centering
  \includegraphics[width=0.9\linewidth]{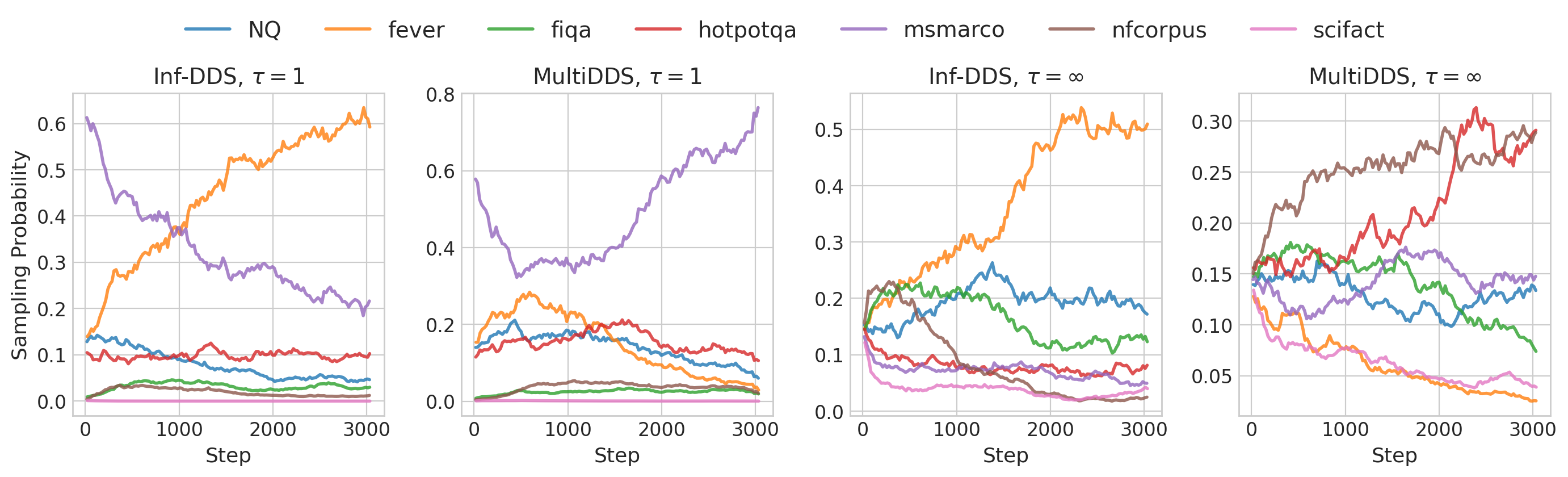}
\caption{Sampling probability trajectories of MultiDDS and \ouralgo{} with varying initialization temperatures during joint optimization on the Dbpedia development set. Since DBpedia is not present in the training set, we see FEVER being upsampled more by \ouralgo{}, which should be true since DBpedia and FEVER are very closely related datasets \cite{DBLP:journals/corr/abs-2104-08663}.}
  \label{fig:plotdbpedia}
\end{figure*}

\begin{figure*}[ht]
  \centering
  \includegraphics[width=\linewidth]{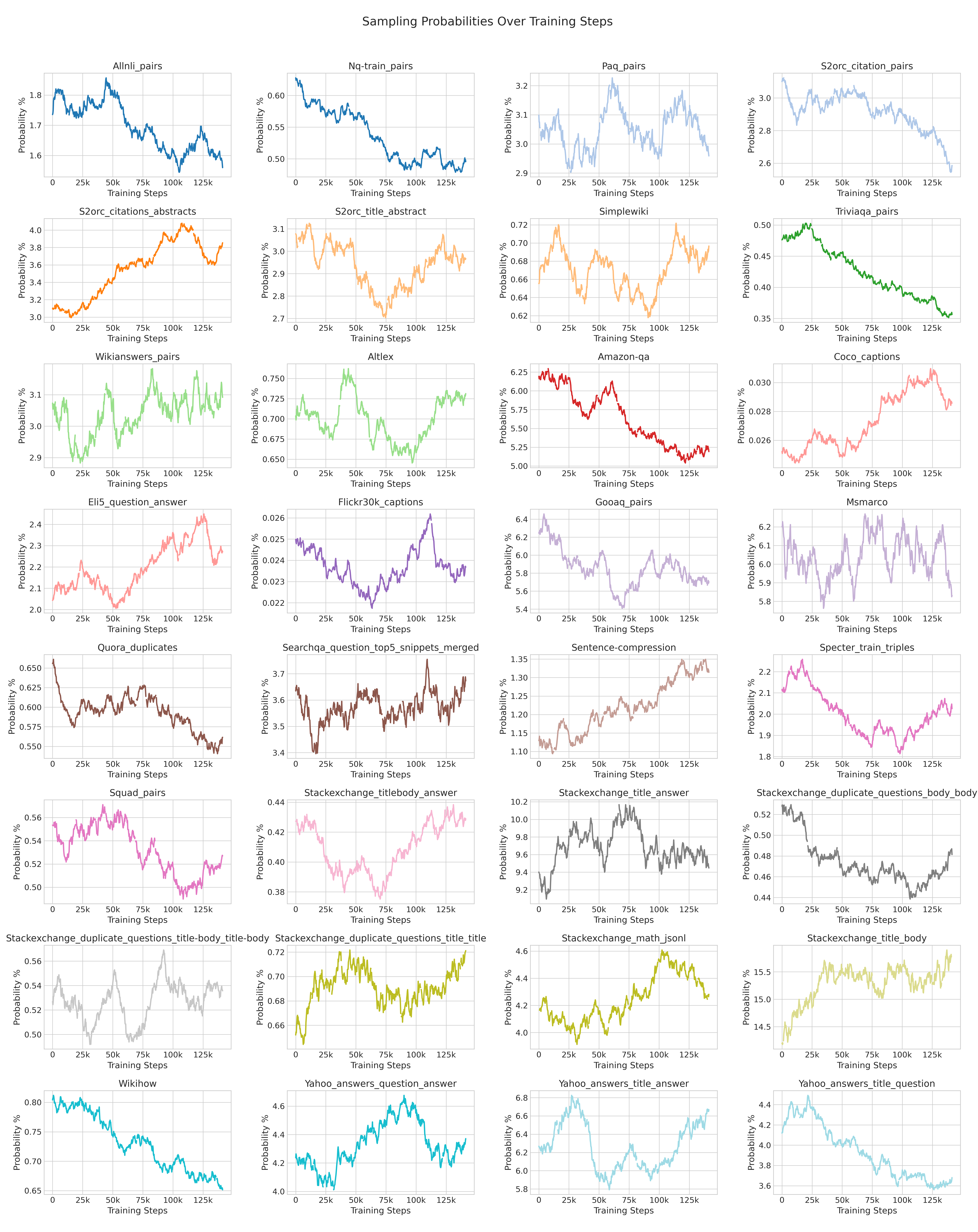}
\caption{Sampling probability trajectories of \ouralgo{} with \emph{Expert} initialization during training of \emph{all-MiniLM-L6-v2} on Sentence Transformers data. Optimization is done jointly on the BEIR-5 development set (DBpedia, FEVER, FiQA, HotpotQA, and Quora).}
  \label{fig:plotinfdds-st}
\end{figure*}

\begin{figure*}[ht]
  \centering
  \includegraphics[width=\linewidth]{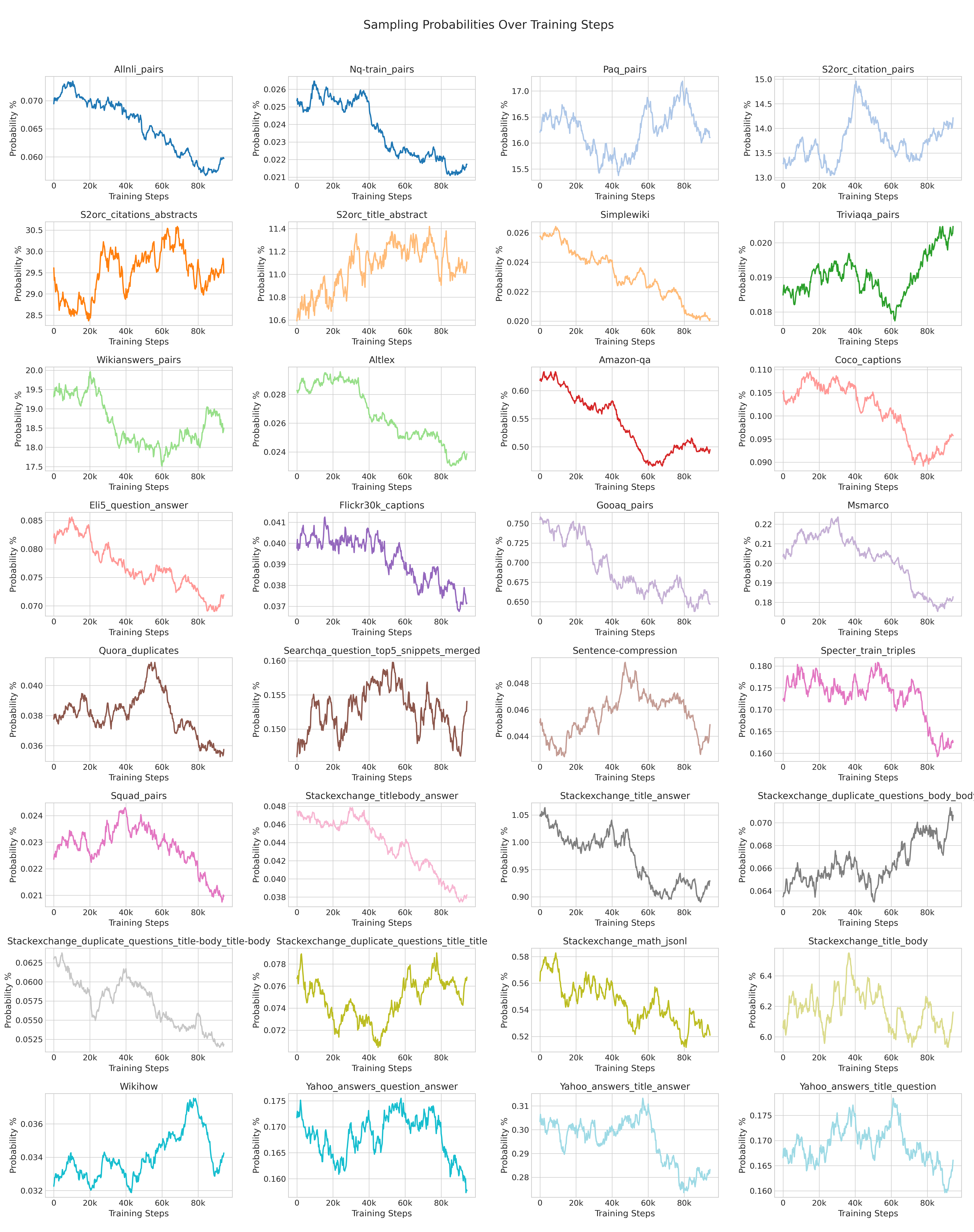}
\caption{Sampling probability trajectories of \ouralgo{} with $\tau=1$ initialization during training of \emph{all-MiniLM-L6-v2} on Sentence Transformers data. Optimization is done jointly on the BEIR-5 development set (DBpedia, FEVER, FiQA, HotpotQA, and Quora).}
  \label{fig:plotinfdds-st-l}
\end{figure*}

\begin{figure*}[ht]
  \centering
  \includegraphics[width=\linewidth]{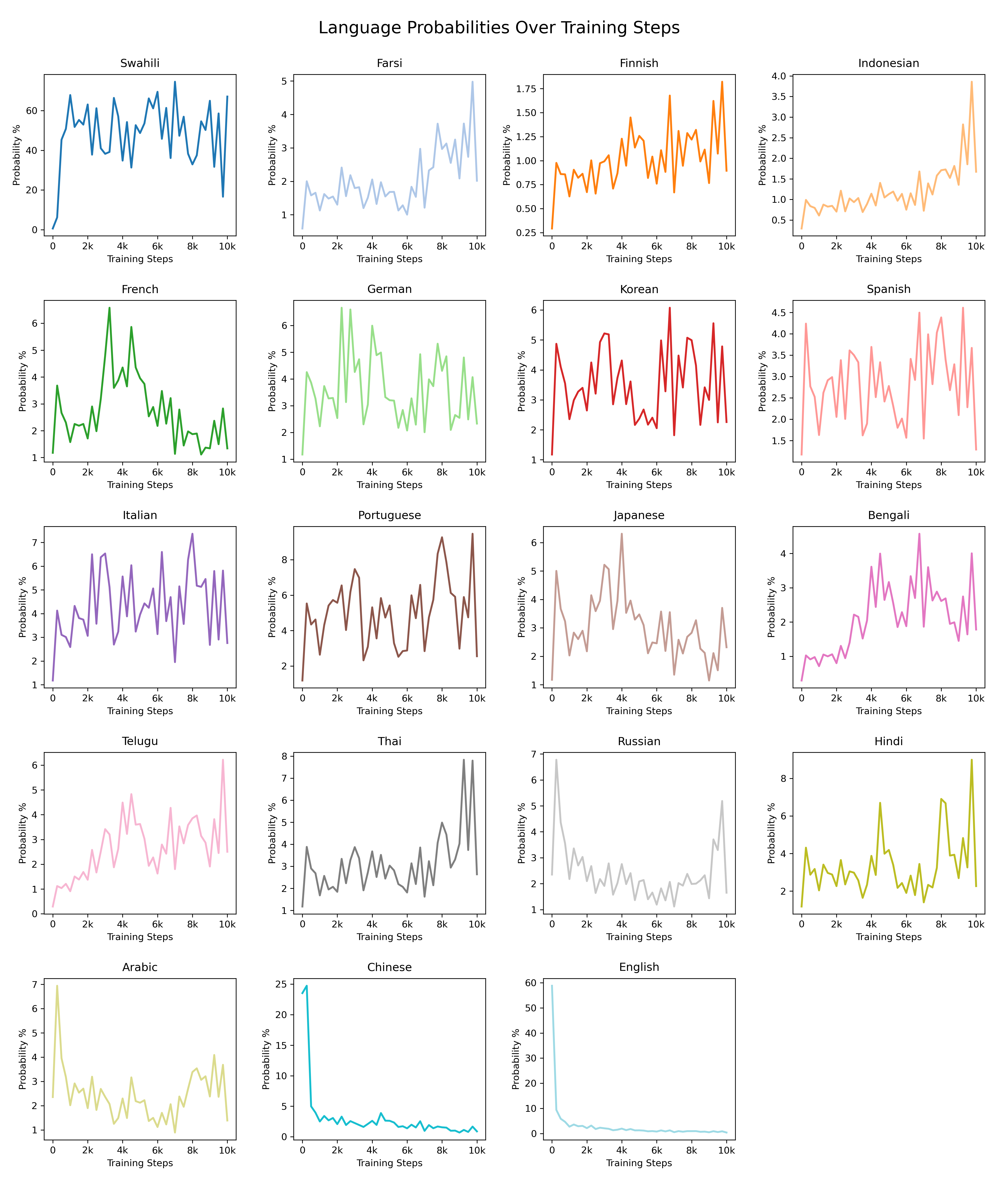}
\caption{Sampling probability trajectories of \ouralgo{} during training of \emph{bge-m3-dense} while jointly optimizing for MLDR-13 dev sets.}
  \label{fig:plotinfdds-mldr13}
\end{figure*}



\end{document}